\documentclass{aa}  

\usepackage{graphicx}
\usepackage{txfonts}
\usepackage{hyperref}
\usepackage{pdflscape}
\usepackage{textcomp}

\hypersetup{colorlinks=true, linkcolor=blue, citecolor=blue, filecolor=blue, urlcolor=blue, pdftitle=, pdfauthor=victor , pdfsubject=, pdfkeywords=}

\begin{document}
                
        \title{In-flight photometry extraction of PLATO targets}
        
        \subtitle{Optimal apertures for detecting extrasolar planets}
        
        \author{
                V. Marchiori \inst{1,2}
                \and
                R. Samadi \inst{1}
                \and
                F. Fialho \inst{2}
                \and
                C. Paproth \inst{3}
                \and
                A. Santerne \inst{4}
                \and
                M. Pertenais \inst{3}
                \and
                A. B{\"{o}}rner \inst{3}
                \and
                J. Cabrera \inst{5}
                \and \\
                A. Monsky \inst{6}
                \and
                N. Kutrowski \inst{7}   
        }
        
        \institute{
                Laboratoire d'Etudes Spatiales et d'Instrumentation en Astrophysique, Observatoire de Paris, Universit{\'{e}} PSL, CNRS, 5 Pl. Jules Janssen, 92190 Meudon, France\\
                \email{victor.marchiori@obspm.fr}
                \and
                Escola Polit{\'{e}}cnica -- Departamento de Engenharia de Telecomunica{\c{c}}{\~{o}}es e Controle, Universidade de S{\~{a}}o Paulo,
                Av. Prof. Luciano Gualberto, trav. 3, n. 158, 05508-010 S{\~{a}}o Paulo, Brazil
                \and
                Deutsches Zentrum f{\"{u}}r Luft- und Raumfahrt (DLR), Institut f{\"{u}}r Optische Sensorsysteme, Rutherfordstra{\ss}e 2, 12489 Berlin-Adlershof, Germany
                \and
                Aix Marseille Universit{\'{e}}, CNRS, CNES, LAM, Marseille, France
                \and
                Deutsches Zentrum f{\"{u}}r Luft- und Raumfahrt (DLR), Institut f{\"{u}}r Planetenforschung, Rutherfordstra{\ss}e 2, 12489 Berlin-Adlershof, Germany
                \and
                OHB System AG, Universit{\"{a}}tsallee 27-29, 28359 Bremen, Germany
                \and
                Thales Alenia Space, 5 Allée des Gabians, 06150 Cannes, France
        }
        
        \date{Received February 14, 2019; accepted May 14, 2019}
        
        
        \abstract
        {The ESA PLATO space mission is devoted to unveiling and characterizing new extrasolar planets and their host stars. This mission will encompass a very large (>2,100 $\mathrm{deg^2}$) field of view, granting it the potential to survey up to one million stars depending on the final observation strategy. The telemetry budget of the spacecraft cannot handle transmitting individual images for such a huge stellar sample at the right cadence, so the development of an appropriate strategy to perform on-board data reduction is mandatory.} 
        {We employ mask-based (aperture) photometry to produce stellar light curves in flight. Our aim is thus to find the mask model that optimizes the scientific performance of the reduced data.}
        {We considered three distinct aperture models: binary mask, weighted Gaussian mask, and weighted gradient mask giving lowest noise-to-signal ratio, computed through a novel direct method. Each model was tested on synthetic images generated for 50,000 potential PLATO targets. We extracted the stellar population from the {\it Gaia} DR2 catalogue. An innovative criterion was adopted for choosing between different mask models. We designated as optimal the model providing the best compromise between sensitivity to detect true and false planet transits. We  determined the optimal model based on simulated noise-to-signal ratio and frequency of threshold crossing events.}
        {Our results show that, although the binary mask statistically presents a few percent higher noise-to-signal ratio compared to weighted masks, both strategies have very similar efficiency in detecting legitimate planet transits. When it comes to avoiding spurious signals from contaminant stars however the binary mask statistically collects considerably less contaminant flux than weighted masks, thereby allowing the former to deliver up to $\sim$30\% less false transit signatures at $7.1\sigma$ detection threshold.}
        {Our proposed approach for choosing apertures has been proven to be decisive for the determination of a mask model capable to provide near maximum planet yield and substantially reduced occurrence of false positives for the PLATO mission. Overall, this work constitutes an important step in the design of both on-board and on-ground science data processing pipelines.}
        
        \keywords{
                instrumentation: photometers -- planets and satellites: detection -- techniques: photometric -- methods: numerical -- catalogs -- zodiacal dust
        }
        
\maketitle


\section{Introduction}  \label{sec:introduction}
PLAnetary Transits and Oscillations of stars (PLATO)\footnote{\url{https://www.cosmos.esa.int/web/plato}} \cite[]{Rauer2014} is a space mission from the European Space Agency (ESA) whose science objective is to discover and characterize new extrasolar planets and their host stars. Expected to be launched by end 2026, this mission will focus on finding photometric transit signatures of Earth-like planets orbiting the habitable zone of main-sequence Sun-like stars. Thanks to its very large field of view ($\sim$2,132 $\mathrm{deg^2}$) covered by multiple (6 to 24) telescopes, PLATO will be able to extract long duration (few months to several years) photometry from a significantly large sample of bright targets ($V<11$) at very high photometric precision ($\sim50 \, \mathrm{ppm \, hr^{1/2}}$). The resulting scientific data are expected to provide stellar ages with accuracy as low as 10\% and radii of Earth-like planets with accuracy as low as 3\% (\cite{ESA2017_RBD} see also \cite{Goupil2017}).

The PLATO data processing pipeline is a critical component of the payload, which is composed of multiple ground- and flight-based algorithms. These are necessary to convert the raw data acquired by the instrument, which inevitably carries unwanted systematic disturbances, into scientifically exploitable light curves. Typical examples of systematic errors are the long-term star position drift, pointing error due to satellite jitter, charge transfer inefficiency (CTI) from the detectors, pixel saturation, outliers, and sky background. To work around these errors, extensive studies have been carried out focussed on the definition of data processing algorithms. These studies include the development of photometry extraction methods which are key for the success of the mission and motivate the present work.

The PLATO photometer will be capable to produce light curves for up to one million stars, depending on the final observation strategy. In contrast, transmitting individual images for each target, at sufficiently short cadence\footnote{Based on mission science requirements, PLATO light curves will be sampled at either 25, 50, or 600 seconds \cite[see][]{ESA2017_RBD}.} for further ground-based processing requires prohibitive telemetry resources. Hence, for a substantial fraction of the targets, an appropriate data reduction strategy (prior to data compression) needs to be executed. In that case, the most suitable encountered solution consists in producing their light curves on board, in a similar way as performed for the targets of Convection, Rotation and planetary Transits (CoRoT) \cite[]{Auvergne2009} Space Telescope. By doing so, the spacecraft transmits data packages to the ground segment containing a single flux value per cadence for each star rather than multiple flux values from several pixels. Within the mission design of PLATO, the group of targets whose light curves will be produced on board are part of a stellar sample called P5. Considering a scenario of two long pointing observations, this set represents more than 245,000 F5 to late-K spectral class dwarf and sub-giant stars with $V$ magnitude ranging from 8 to 13; it was idealized to generate large statistical information on planet occurrence rate and systems evolution. For all other stellar samples, which are primarily composed of the brightest targets \cite[more details in][]{ESA2017_RBD}, the photometry will be extracted on the ground from individual images, thereby following the same principle as that of {\it Kepler} Space Telescope \cite[]{Borucki2010} and Transiting Exoplanet Survey Satellite (TESS) \cite[]{Ricker2014} targets.

In view of its acknowledged high performance and straightforward implementation, mask-based (aperture) was adopted as in-flight photometry extraction method to be implemented in the PLATO data processing pipeline. In such technique, each light curve sample is generated by integrating the target flux over a limited number of pixels, which shall be appropriately selected to maximize the scientific exploitability of the resulting time-series light curve. In this context, the present work unfolds the development carried out for defining the optimal collection of pixels for extracting photometry from non-saturated stars in the P5 sample.

There is a noteworthy number of publications on the theme of photometric masks. Among the oldest, we put some emphasis on the work of \cite{Howell1989}, in which the idea of a growth curve (signal-to-noise ratio as a function of aperture radius) for point-source observations is presented; on the stellar photometry package DAOPHOT\footnote{\url{http://www.star.bris.ac.uk/~mbt/daophot/}.} from \cite{Stetson1987}, which is still widely used today; and on the solution proposed by \cite{Naylor1998}, which consists of employing weighted masks for imaging photometry, providing improved noise-to-signal ratio (NSR) performance compared to binary masks. Later on, and orientated to planet transit finding and asteroseismology, \cite{Llebaria2006} and \cite{Bryson2010} developed strategies to compute optimized binary masks\footnote{In the context of {\it Kepler's} data processing pipeline, such an aperture is referred to as simple aperture photometry. It was primarily designed to minimize noise for maximum transit detection sensitivity and as input for determining a halo of pixels to be downlinked along with the aperture pixels.} for extracting light curves from CoRoT and {\it Kepler} targets, respectively. More recently, \cite{Smith2016} proposed a new method to assign apertures for {\it Kepler} targets, focussed on planet detection and mitigation of systematic errors, through an optimization scheme based on NSR and Combined Differential Photometric Precision (CDPP)\footnote{Roughly speaking, CDPP is an estimate of how well a transit-like signal can be detected \cite[]{Smith2016}.} \cite[]{Jenkins2010b}. As described in {\it Kepler's} Data Processing Handbook\footnote{\url{https://archive.stsci.edu/kepler/manuals/}}, this method is implemented within the photometry analysis component of {\it Kepler's} science pipeline. Alongside, \cite{Aigrain2015} and \cite{Lund2015} provided techniques for mask pixel selection for {\it Kepler K2} targets. The former proposes circular apertures, which has satisfactory performance for sufficiently bright targets and is relatively robust to systematic errors. The latter uses clustering of pixels, which best fits the flux distribution of the targets, being therefore more suitable for dense fields. A modified version of this method is employed in \cite{Handberg2016} for reducing the data of {\it Kepler K2} targets from campaigns 0 to 4. Besides, it is also considered as one of the possible solutions for extracting light curves from TESS targets \cite[]{Lund2017}.

In this paper, we are evidently interested in solutions that are better suited for both exoplanet search and asteroseismology, which brings thus our attention to those that were developed for the space missions CoRoT, {\it Kepler,} and TESS. Considering these three examples, we notice that the notion of optimal aperture is employed to distinguish apertures that minimize NSR or some noise-related metric such as CDPP. That is, of course, a reasonable way to proceed because the sensitivity at which a planet transit can be found in a light curve, for instance, is strongly correlated to its noise level. On the other hand, the higher the ease in identifying a transit-like signal, either because of sufficiently low NSR or CDPP, the higher the probability that a background object in the scene generates a threshold crossing event (TCE)\footnote{This concept was created in the context of the {\it Kepler} science pipeline and designates a statistically significant transit-like signature marked for further data validation \cite[e.g. see][]{Twicken2018}.}. This background object could be, for example, a stellar eclipsing binary (EB) mimicking
a true planet transit. Background false positives may be efficiently identified in certain cases when, besides the light curves, the corresponding pixel data is also available, as demonstrated by \cite{Bryson2013}; however, most of the stars in P5 unfortunately lack that extra information\footnote{For an observation scenario covering two long pointing fields, the telemetry budget dedicated to the P5 sample includes, in addition to the light curves, more than 9,000 imagettes -- at 25 seconds cadence -- and centre of brightnesses (COB) for 5\% of the targets \cite[]{ESA2017_RBD}.} because of telemetry constraints already mentioned. Under such an unfavourable scenario, conceiving photometric masks based uniquely on how well a transit-like signal can be detected, paying no attention to potential false positives may not be the best strategy. To verify the consistence of this hypothesis, we introduce in this paper two science metrics that allow us to directly quantify the sensitivity of an aperture in detecting true and false\footnote{In this paper, we address the occurrence of false planet transits caused by background eclipsing objects, in particular EBs.} planet transits. Then we determine whether or not the best compromise between these two parameters is obtained from apertures having overall lower NSR.

\begin{figure}
        \begin{center}
                \includegraphics[height=7.5cm, width=\hsize]{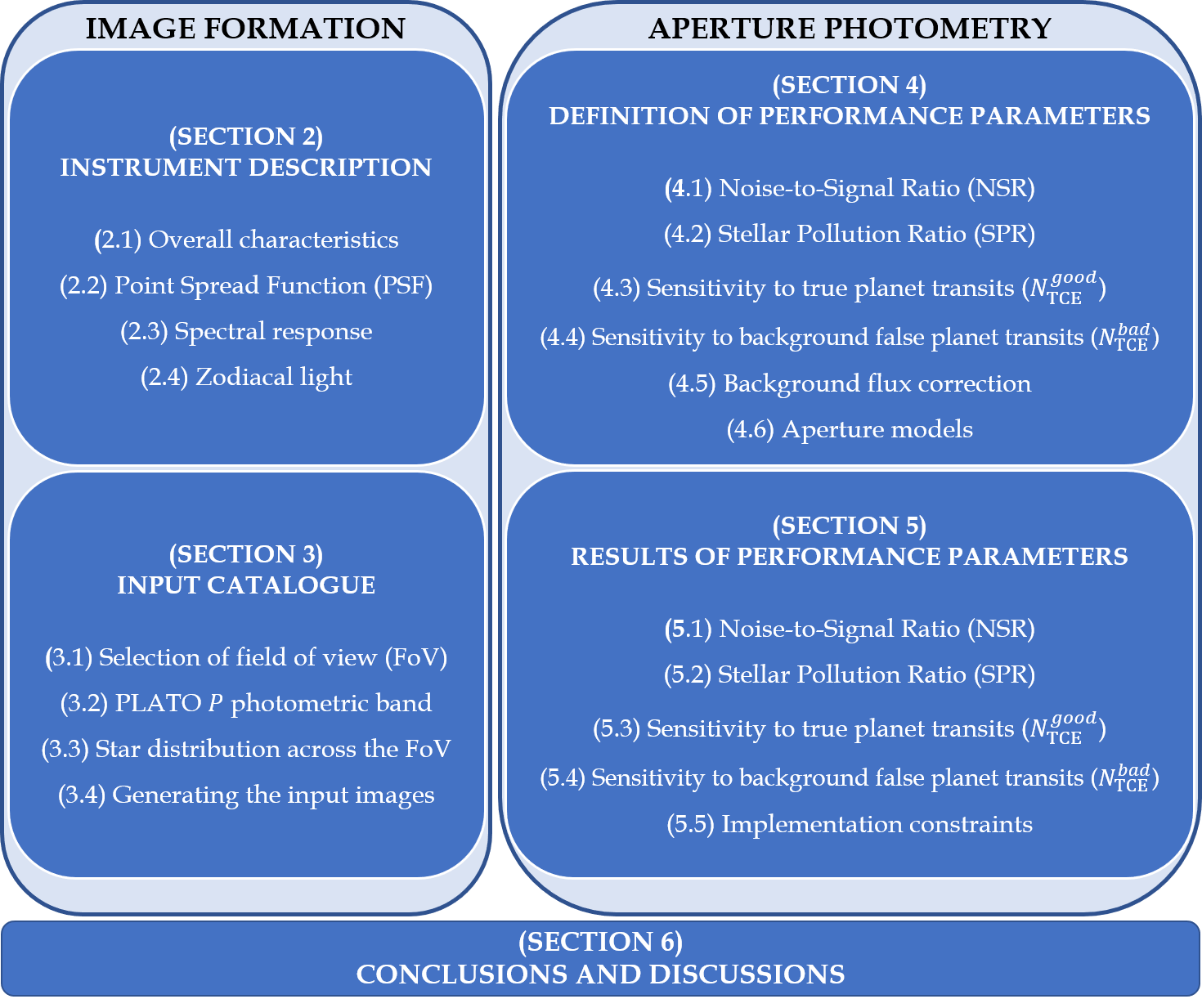}
                \caption{Overview of paper content.}
                \label{fig:overview}
        \end{center}
\end{figure}

This paper is organized as follows (see Fig.~\ref{fig:overview}). Section~\ref{sec:instrument} describes the main payload characteristics, including instrument  point spread function (PSF), spectral response, and noise. Also, an expression is derived to provide an estimation on the intensities of zodiacal light entering each PLATO telescope. Section~\ref{sec:input_catalogue} gives details on the extracted data from the adopted input catalogue ({\it Gaia} DR2). That information is used to build synthetic input images, called imagettes, to characterize the performance of aperture photometry. A synthetic PLATO $P$ photometric passband, calibrated on the VEGAMAG system, is derived to avoid the inconvenience of having colour dependency when estimating stellar fluxes -- at detector level -- from visual magnitudes. Colour relationships with Johnson's $V$ and {\it Gaia} $G$ magnitudes are provided. Section~\ref{sec:aperture_photometry} describes the methodology applied to find the optimal aperture model to extract photometry from stars in P5 sample. Three models are tested, including a novel direct method for computing a weighted aperture providing global lowest NSR. Section~\ref{sec:results} shows comparative results between all aperture models with respect to their sensitivity in detecting true and false planet transits. Lastly, Section~\ref{sec:conclusion} concludes with discussions on the presented results.

\section{The instrument} \label{sec:instrument}

\subsection{Overall characteristics} \label{sec:overall}
\begin{figure}
        \centering
        \includegraphics[height=4cm, width=4cm]{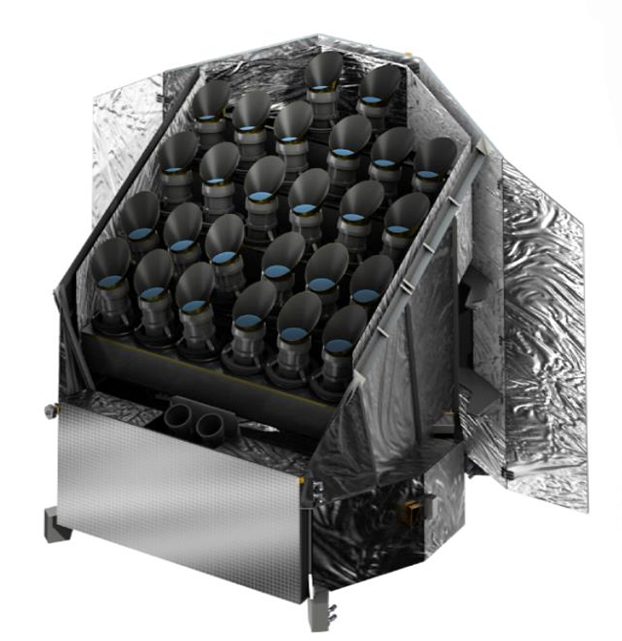}
        \includegraphics[height=4cm, width=4cm]{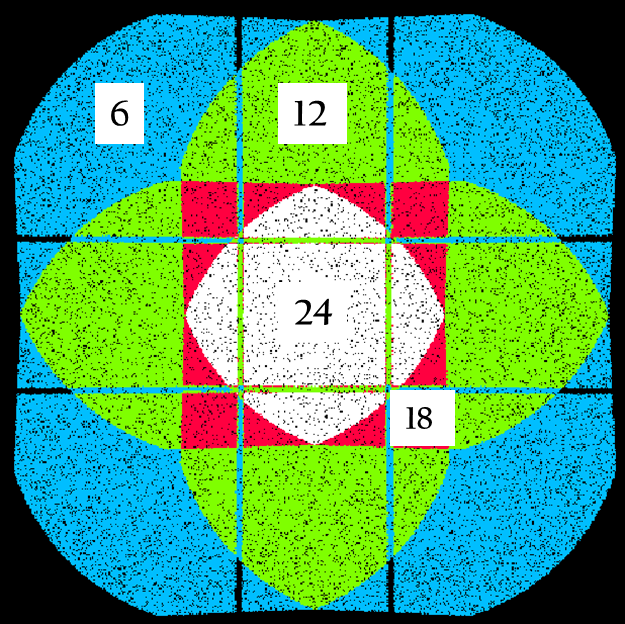}
        \caption{\textbf{Left}: Representation of the PLATO spacecraft with 24+2 telescopes. Credit: OHB-System AG. \textbf{Right}: Layout of the resulting field of view obtained by grouping the normal telescopes into a $4\times6$ overlapping configuration. The colour code indicates the number of telescopes covering the corresponding fractional areas (Table~\ref{table:instrument_data}): 24 (white), 18 (red), 12 (green), and 6 (cyan). Credit: The PLATO Mission Consortium.}
        \label{fig:multitelescope_scheme}
\end{figure}

The PLATO payload relies on an innovative multi-telescope concept consisting of 26 small aperture (12 cm pupil diameter) and wide circular field of view ($\sim$1,037 $\mathrm{deg^2}$) telescopes mounted in a single optical bench. Each telescope is composed of an optical unit (TOU), a focal plane assembly  holding the detectors, and a front-end electronics (FEE) unit. The whole set is divided into 4 groups of 6 telescopes (herein called normal telescopes or N-CAM) dedicated to the core science and 1 group of 2 telescopes (herein called fast telescopes or F-CAM) used as fine guidance sensors by the attitude and orbit control system. The normal telescope assembly results in a overlapped field of view arrangement (see Fig.~\ref{fig:multitelescope_scheme}), allowing them to cover a total sky extent of about 2,132 $\mathrm{deg^2}$, which represents almost 20 times the active field of the {\it Kepler} instrument. The N-CAM and F-CAM designs are essentially the same, except for their distinct readout cadence (25 and 2.5 seconds, respectively) and operating mode (full-frame and frame-transfer, respectively). In addition, each of the two F-CAM includes a bandpass filter (one bluish and the other reddish) for measuring stellar flux in two distinct wavelength bands. Table~\ref{table:instrument_data} gives an overview of the main payload characteristics based on \cite{ESA2017_RBD}.
\begin{table}
        \caption[]{Summary of main payload characteristics.}
        \label{table:instrument_data}
        \centering
        \begin{tabular}{cc}
                \hline
                \noalign{\smallskip}
                Description & Value \\ [0.5ex]
                \hline 
                \noalign{\smallskip \smallskip}
                \small  Optics  & \small (24+2) telescopes with  \\
                                            & axisymmetric dioptric design \\
                \noalign{\smallskip \smallskip}
                TOU Spectral range & $\mathrm{500 - 1,000 \, nm}$  \\
                \noalign{\smallskip \smallskip}
                Pupil diameter & $\mathrm{12 \, cm}$ \\
                (per telescope) & \\
                \noalign{\smallskip \smallskip}
                Detector & back-illuminated \\
                         & Teledyne-e2v CCD 270 \\
                \noalign{\smallskip \smallskip}
                N-CAM focal plane & 4 full-frame CCDs \\ 
                                        & (4,510 $\times$ 4,510 pixels each) \\              
                \noalign{\smallskip \smallskip}
                F-CAM focal plane & 4 frame-transfer CCDs \\ 
                                          & (4,510 $\times$ 2,255 pixels each) \\        
                \noalign{\smallskip \smallskip}
                Pixel length & $\mathrm{18 \, \mu m}$ \\
                \noalign{\smallskip \smallskip}
                On-axis plate scale & 15 arcsec \\
                (pixel field of view) & \\
                \noalign{\smallskip \smallskip}
                Quantization noise & $\mathrm{\sim 7.2 \, e^- rms/px}$ \\               
                \noalign{\smallskip \smallskip}
                Readout noise (CCD+FEE) & $\mathrm{\sim 50.2 \, e^- rms/px}$ \\
                at beginning of life & \\
                \noalign{\smallskip \smallskip}
                Detector smearing noise & $\mathrm{\sim 45 \, e^- /px/s}$ \\      
                \noalign{\smallskip \smallskip}
                Detector dark current noise & $\mathrm{\sim 4.5 \, e^- /px/s}$ \\      
                \noalign{\smallskip \smallskip}
                N-CAM cadence & $\mathrm{25 \, s}$ \\
                \noalign{\smallskip \smallskip}
                N-CAM exposure time & $\mathrm{21 \, s}$ \\
                \noalign{\smallskip \smallskip}
                N-CAM readout time & $\mathrm{4 \, s}$ \\
                \noalign{\smallskip \smallskip}
                F-CAM cadence & $\mathrm{2.5 \, s}$ \\  
                \noalign{\smallskip \smallskip}
                N-CAM field of view & $\mathrm{\sim 1,037 \, deg^2}$ (circular) \\
                \noalign{\smallskip \smallskip}
                F-CAM field of view & $\mathrm{\sim 619 \, deg^2}$ \\
                \noalign{\smallskip \smallskip}
                Full field of view & $\mathrm{\sim 2,132 \, deg^2}$ \\
                \noalign{\smallskip \smallskip}
                Fractional field of view & $\mathrm{294 \, deg^2}$ (24 telescopes)  \\
                                                                 & $\mathrm{171 \, deg^2}$ (18 telescopes) \\
                                                                 & $\mathrm{796 \, deg^2}$ (12 telescopes) \\
                                                                 & $\mathrm{871 \, deg^2}$ (6 telescopes) \\
                \hline                                    
        \end{tabular}
\end{table}

\subsection{Point spread function} \label{sec:psf}
\begin{figure}
        \centering
        \includegraphics[height=5.5cm, width=\hsize]{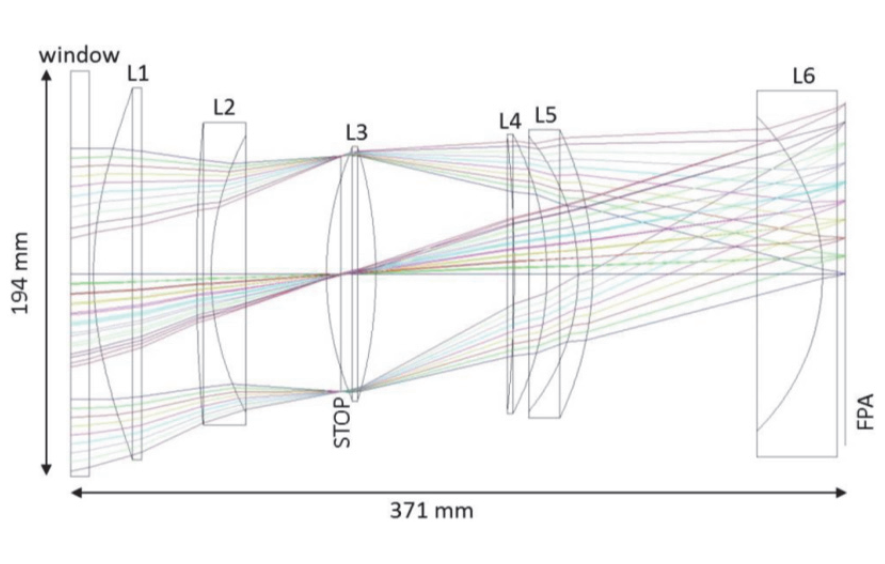}
        \caption{Baseline optical layout of each PLATO telescope. Credit: The PLATO Mission Consortium.}
        \label{fig:optical_layout}
\end{figure}

Starlight reaching the focal plane of PLATO cameras will inevitably suffer from distortions caused by both optics and detectors, causing this signal to be non-homogeneously spread out over several pixels. The physical model describing such effects is the PSF, from which we can determine -- at subpixel level -- how stellar signals are distributed over the pixels of the detector. In this work, we used synthetic optical PSF models obtained from the baseline telescope optical layout (Fig.~\ref{fig:optical_layout}) simulated on ZEMAX\textsuperscript{\textregistered} software. Estimated assembly errors such as lens misalignment and focal plane defocus are included.

\begin{figure}
        \centering
        \includegraphics[height=14.5cm, width=\hsize]{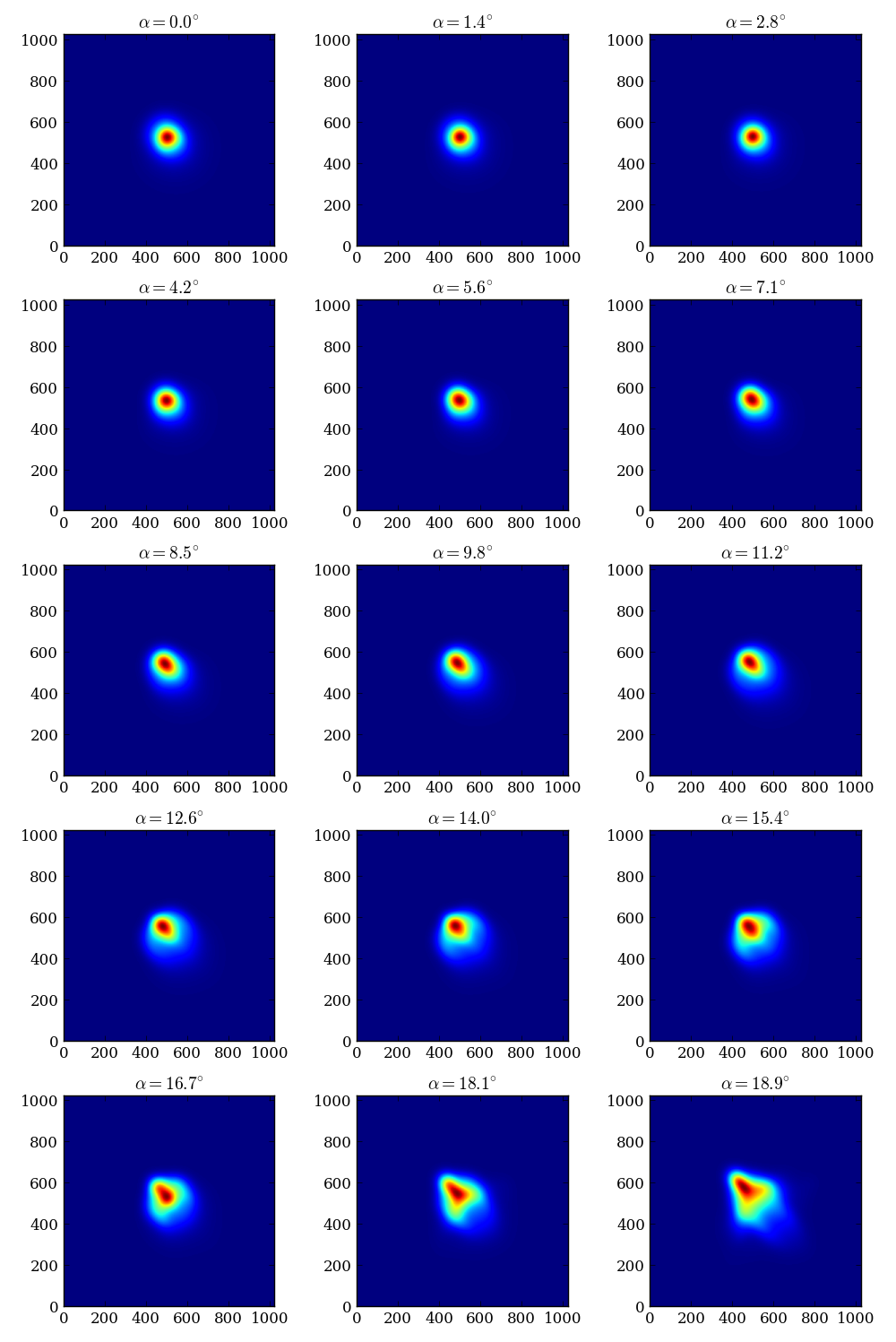}
        \caption{Simulated PSF shapes of PLATO telescopes (1/128 pixel resolution) as a function of the angular position, $\alpha$, in the sky of a source at $-45^{\circ}$ azimuth. Angular positions range from $\alpha=0^{\circ}$ (centre) to  $\alpha=18.9^{\circ}$ (edge) of the camera field of view. Each optical PSF is convolved to a Gaussian diffusion kernel with standard deviation of 0.2 pixel to simulate the CCD behaviour. Each image above corresponds to a CCD surface of $8\times8$ pixels, which is enough to encompass $\sim$99.99\% of the total PSF energy.}
        \label{fig:all_psfs}
\end{figure}

Beyond optics, the detectors also degrade the spatial resolution of stellar images through charge disturbances processes such as CTI \cite[]{Short2013,Massey2014}, ``brighter-fatter'' \cite[]{Guyonnet2015}, and diffusion \cite[]{Widenhorn2010}. Several tests are being carried out by ESA to characterize such effects for the charge coupled devices (CCD) of PLATO cameras, so at the present date no formal specifications for the corresponding parameters are available. However, the optical PSFs alone are known to be a non-realistic final representation of the star signals. Therefore, to obtain a first order approximation of the real physics behind the PSF enlargement taking place at PLATO detectors with respect to the diffusion, the optical PSFs are convolved to a Gaussian kernel with a standard deviation of 0.2 pixel. The resulting simulated PSF models are shown in Fig.~\ref{fig:all_psfs} for 15 angular positions, $\alpha$, within the field of view of one camera. In this paper, PSF shape variations due to target colour are assumed to be of second order and are thus ignored.

\begin{figure}
        \centering
        \includegraphics[height=6.0cm, width=\hsize]{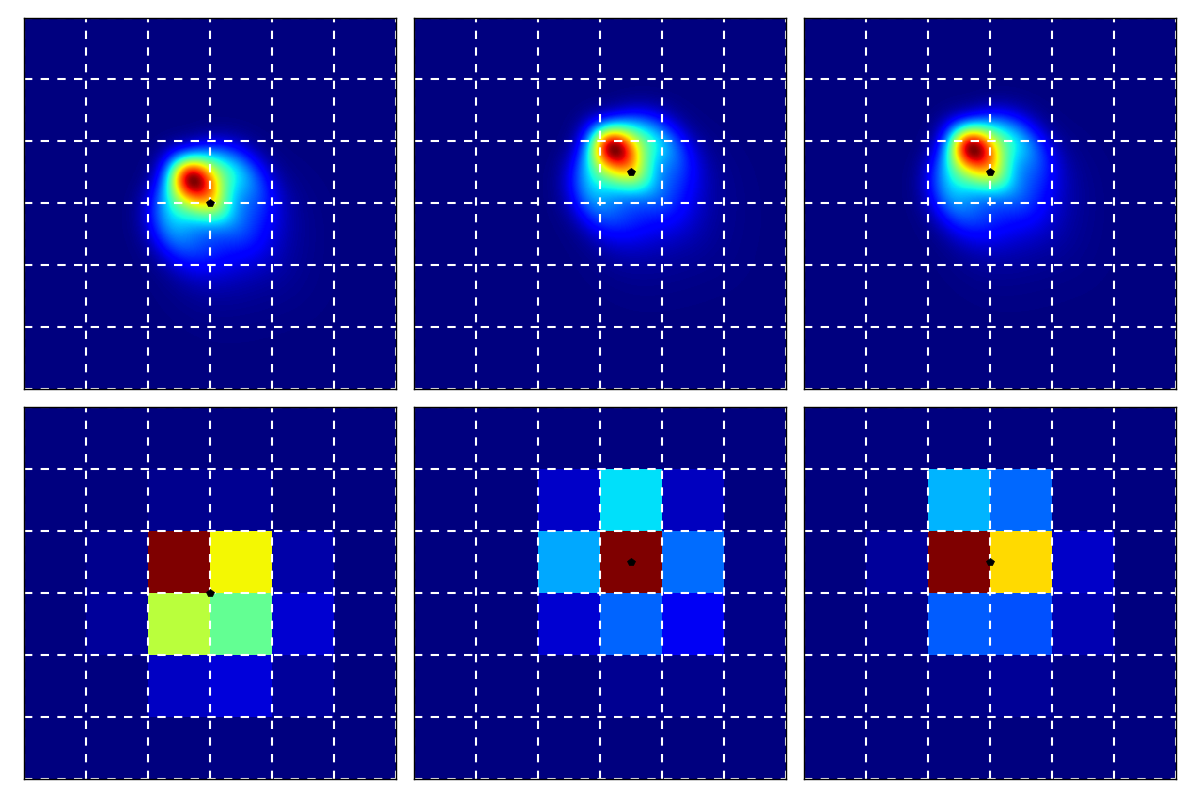}
        \caption{Energy distribution of PSF across the pixels for three distinct intra-pixel target barycentre locations (black dots): at pixel corner (left), at pixel centre (middle), and at the border of two adjacent pixels (right). Dashed white lines represent pixel borders. \textbf{Top}: High resolution PSF at $\alpha=14^{\circ}$. \textbf{Bottom}: Corresponding low resolution PSF.}
        \label{fig:psf_imagette}
\end{figure}

To reduce the overlap of multiple stellar signals and increase photometric precision, PLATO cameras are primarily designed to ensure that about 77\% of the PSF flux is enclosed, on average, within $\mathrm{\sim 2.5 \times 2.5 \, pixels}$ across the field of view, or 99\% within $\mathrm{\sim 5 \times 5 \, pixels}$. As a consequence, the size of the pixels are relatively large compared to that of the PSF, making the distribution of energy from stars very sensitive to their barycentre location within a pixel (see Fig.~\ref{fig:psf_imagette}).

During and after launch, the space environment unavoidably causes overall changes in the instrument response that cannot always be accurately predicted, including variations in the PSF model. Nevertheless, accurate knowledge of the PSFs is imperative for proper correction of systematic errors in the light curves and computing the photometric apertures, so a strategy for reconstructing the PSFs is needed. As the individual raw images downlinked from the spacecraft cannot describe the distribution of stellar flux on the detectors with sufficient resolution, high resolution PSFs such as those in Fig.~\ref{fig:all_psfs} will be reconstructed on the ground from micro-scanning sessions \cite[]{Samadi2019}. This process, which will occur every three months during instrument calibration phases, basically consists of acquiring a series of raw images from subpixel displacements following an Archimedean spiral. Then, inverse methods are employed to reconstruct the PSFs from the lower resolution micro-scanning images.

\subsection{Spectral response and vignetting} \label{sec:spectral_response}
\begin{figure}
        \centering
        \includegraphics[height=6.5cm, width=\hsize]{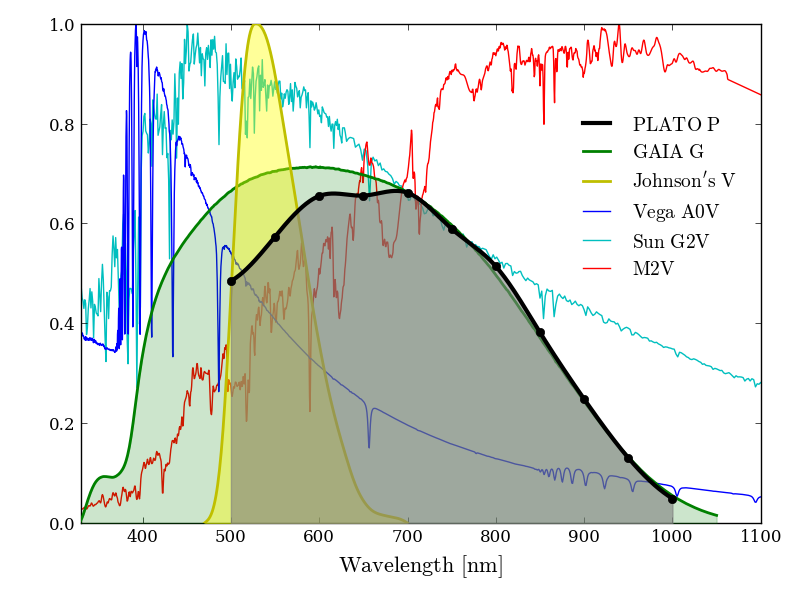}
        \caption{\textbf{Black}: Preliminary spectral response of PLATO N-CAM at beginning of life. Values are currently known at the black dots; these are crossed by a cubic spline interpolation curve. \textbf{Green}: {\it Gaia} $G$ band. \textbf{Yellow}: Johnson's V filter. \textbf{Blue}: Vega (A0V) normalized spectrum. \textbf{Cyan}: Sun (G2V) normalized spectrum. \textbf{Red}: Normalized spectrum of M2V-type star.}
        \label{fig:plato_throughput}
\end{figure}

The spectral response of a photometer represents its efficiency in converting incident photons into effective counts of electrons at detector level. It is derived from the combined effect of optical transmission and CCD quantum efficiency. In parallel, another parameter impacting instrument efficiency is vignetting, which consists of an inherent optical feature that causes attenuation of image brightness. Such an effect increases non-linearly as the angular position, $\alpha$, of the source augments with respect to the optical axis ($\alpha=0$) of the instrument. A preliminary spectral response curve of the PLATO cameras is presented in Fig.~\ref{fig:plato_throughput}, alongside the {\it Gaia} $G$ passband\footnote{\url{https://www.cosmos.esa.int/web/gaia/auxiliary-data}}, Johnson's V filter from \cite{Bessell1990}, Vega A0V Kurucz template ($\mathrm{\texttt{alpha\_lyr\_stis\_008}}$) from the CALSPEC\footnote{\url{http://www.stsci.edu/hst/observatory/crds/calspec.html}} database, E-490\footnote{\url{https://www.nrel.gov/grid/data-tools.html}} reference solar spectrum from the American Society for Testing and Materials (ASTM), and a M2V-type star synthetic spectrum from the Pickles atlas\footnote{\url{http://www.stsci.edu/hst/observatory/crds/pickles_atlas.html}} \cite[]{Pickles1998}. Vignetting intensities, $\mathrm{f_{vig}}$, are given in Table~\ref{table:plato_vignetting} as a function of the off-axis angle, $\alpha$, of the target.

\begin{table}
        \caption[]{Combined natural and mechanical obscuration vignetting, $\mathrm{f_{vig}}$, as a function of the off-axis angle, $\alpha$, of the target.}
        \label{table:plato_vignetting}
        \centering
        \begin{tabular}{c c | c c | c c}
                \hline
                \noalign{\smallskip}
                $\alpha$ [deg] & $\mathrm{f_{vig}}$ [\%] & $\alpha$ [deg] & $\mathrm{f_{vig}}$ [\%] & $\alpha$ [deg] & $\mathrm{f_{vig}}$ [\%] \\ [0.5ex]
                \hline 
                \noalign{\smallskip}
                0.000 & 0.00 & 7.053  & 1.51 & 14.001 & 5.85 \\
                \noalign{\smallskip}
                1.414 & 0.06 & 8.454  & 2.16 & 15.370 & 7.03 \\
                \noalign{\smallskip}
                2.827 & 0.24 & 9.850  & 2.93 & 16.730 & 8.53 \\
                \noalign{\smallskip}
                4.238 & 0.55 & 11.241 & 3.80 & 18.081 & 11.58 \\
                \noalign{\smallskip}
                5.647 & 0.97 & 12.625 & 4.78 & 18.887 & 13.69 \\
                \hline                                        
        \end{tabular}
\end{table}

\subsection{Zodiacal light} \label{sec:sky_background}
Scattered sky background light account for significant noise contribution, impacting photometry performance. Hence, we are interested in estimating the amount of diffuse background flux captured by the detectors of each PLATO camera. As the spacecraft will be positioned in L2 orbit (located at approximately $\mathrm{1.01}$ au from the Sun), sky background flux entering its cameras will be dominated by the zodiacal light, i.e. sunlight scattered by interplanetary dust particles agglomerated across the ecliptic plane. Zodiacal light brightness is conventionally expressed in counts of 10th visual magnitude solar-type stars per square degree, also known as $\mathrm{S_{10}}$ unit. By denoting $F_\odot(\lambda)$ as the solar spectrum at 1 au and adopting a corresponding apparent visual magnitude $V_{\mathrm{Sun}}=-26.74$ mag, the $\mathrm{S_{10}}$ unit is formally defined as
\begin{equation}\label{eq:S10}
        \mathrm{S_{10}} = 10^{-0.4 (10 - V_{\mathrm{Sun}})} F_\odot(\lambda) \,\, \mathrm{deg^{-2}} = 6.61 \times 10^{-12} F_\odot(\lambda) \,\, \mathrm{sr^{-1}}. \\
\end{equation}
Tabular data containing zodiacal light measurements in $\mathrm{S_{10}}$ units are available in \cite{Leinert1998}. The published values are valid for an observer located in the vicinity of Earth and at monochromatic wavelength (500 nm). Outside these conditions, a semi-analytical model containing a few correction factors shall be applied. Based on that model we have built an expression (see Table~\ref{table:zl_parameters} for parameters description) to estimate the amount of zodiacal light flux $f^{P}_{\mathrm{ZL}}$ on one N-CAM
\begin{equation}\label{eq:S10_plato}
\begin{split}
        f^{P}_{\mathrm{ZL}} = f_{\mathrm{ZL}} \left( 6.61 \times 10^{-12} \,\, \mathrm{sr^{-1}} \right) \left( h \, c \right)^{-1} \Omega \, \Theta \, \mathrm{f_{L2}} \left( 1 - \mathrm{f_{vig}} \right) \times \\
        \times \int_{\lambda_1}^{\lambda_2} F_\odot(\lambda) \, \mathrm{f_{red}}(\lambda) \, S(\lambda) \, \lambda \, d\lambda. \\
\end{split}
\end{equation}
\begin{table}
        \caption[]{Description of the parameters of Eq.~\ref{eq:S10_plato}.}
        \label{table:zl_parameters}
        \centering
        \resizebox{\columnwidth}{!}{
        \begin{tabular}{c c c c}
                \hline
                \noalign{\smallskip}
                Description & Symbol & Value & Unit \\ [0.5ex]
                \hline 
                \noalign{\smallskip\smallskip}
                Zodiacal light tabulated data & $f_{\mathrm{ZL}}$ & see \cite{Leinert1998} & $\mathrm{S_{10}}$ \\
                \noalign{\smallskip\smallskip}
                Planck's constant & $h$ & $6.63 \times 10^{-34}$ & $\mathrm{J \, s}$ \\
                \noalign{\smallskip\smallskip}
                Speed of light in vacuum & $c$ & $2.99 \times 10^{8}$ & $\mathrm{m\,s^{-1}}$ \\
                \noalign{\smallskip\smallskip}
                Field of view solid angle (per pixel) & $\Omega$ & $4.2 \times 10^{-9}$ & $\mathrm{sr}$ \\
                \noalign{\smallskip\smallskip}
                Entrance pupil surface (per camera) & $\Theta$ & 113.1 & $\mathrm{cm^2}$ \\
                \noalign{\smallskip\smallskip}
                Spectral range of PLATO telescopes & $[\lambda_1, \lambda_2]$ & $[500, 1000]$ & $\mathrm{nm}$ \\
                \noalign{\smallskip\smallskip}
                Sun's spectral irradiance & $F_\odot(\lambda)$ & E-490 spectrum & $\mathrm{W/cm^2/nm}$ \\
                \noalign{\smallskip\smallskip}
                Correction factor for L2 point & $\mathrm{f_{L2}}$ & 0.975 & $\mathrm{adim}$ \\
                \noalign{\smallskip\smallskip}
                Instrument vignetting & $\mathrm{f_{vig}}$ & see Table~\ref{table:plato_vignetting} & $\mathrm{adim}$ \\
                \noalign{\smallskip\smallskip}
                Reddening correction factor & $\mathrm{f_{red}}(\lambda)$ & see \cite{Leinert1998} & $\mathrm{adim}$ \\
                \noalign{\smallskip\smallskip}
                Spectral response of PLATO telescopes & $S(\lambda)$ & see Fig.~\ref{fig:plato_throughput} & $\mathrm{adim}$ \\
                \noalign{\smallskip}
                \hline                                    
        \end{tabular}
        }
\end{table}
Modelling $F_\odot(\lambda)$ with ASTM's E-490 reference solar spectrum (see Fig.~\ref{fig:plato_throughput}) results (expressed in units of $\mathrm{e^- \, px^{-1} \, s^{-1}}$) in
\begin{equation}\label{eq:S10_ccd}
f^{P}_{\mathrm{ZL}} \sim 0.39 \, f_{\mathrm{ZL}} \left( 1 - \mathrm{f_{vig}} \right).
\end{equation}
Excluding vignetting ($f_{\mathrm{vig}} = 0$), $1 \, \mathrm{S_{10}}$ of zodiacal light ($f_{\mathrm{ZL}} = 1$) corresponds thus to about 0.39 $\mathrm{e^- \, px^{-1} \, s^{-1}}$ being generated at the detectors of one PLATO camera. We note that this value might be updated in the future, depending on the final spectral response of the instrument.

\section{Input stellar catalogue}\label{sec:input_catalogue}
An input stellar catalogue is an essential tool for space missions dedicated to asteroseismology and exoplanet searches. Besides its crucial role in field and target selection, it is also noticeably useful for estimating and characterizing the performance of photometry extraction methods prior to mission launch. Indeed, an input catalogue allows us to produce synthetic sky images containing realistic stellar distributions, including their relative positions, apparent magnitudes, effective temperatures, gravities, metallicities, and more. At the present date, a PLATO Input Catalogue (PIC) is being developed based on the ultra-high precision astrometric data from the {\it Gaia} mission \cite[]{GaiaCollaboration2016}. In the future, the PIC might also include information available from other sky surveys such as the Large Synoptic Survey Telescope (LSST) \cite[]{Ivezic2008}, Panoramic Survey Telescope and Rapid Response System (PanSTARSS) \cite[]{Chambers2016}, and SkyMapper \cite[]{Wolf2018}. The PIC will provide abundant and detailed stellar information for optimized target selection vis-à-vis mission science goals. As the PIC was not yet available\footnote{An early version of the PIC is currently available for PLATO consortium members upon request to the PLATO Data Center Board.} by the time that the present work was started, we have adopted the {\it Gaia} data release 2 (DR2) \cite[]{GaiaCollaboration2018b} as our input catalogue, which provides all the information needed for the present work.
\subsection{Observing strategy and input field selection}\label{sec:input_field}
\begin{figure}
        \begin{center}
                \includegraphics[height=4.5cm, width=\hsize]{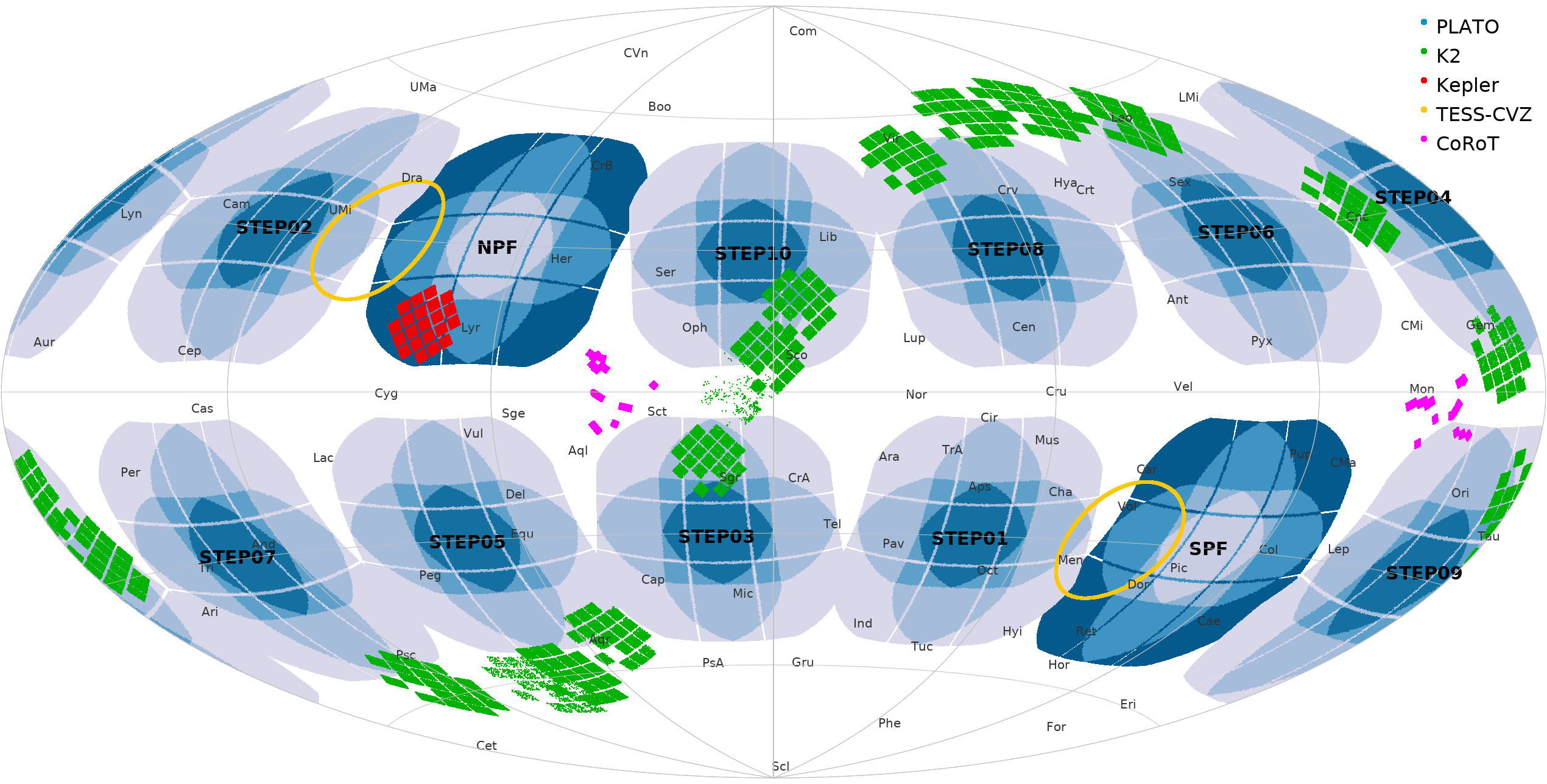}
        \end{center}
        \caption{Sky coverage in Galactic coordinates of PLATO's provisional SPF and NPF long-duration LOP fields, including the possible locations of the short-duration SOP fields (STEP 01-10), to be defined two years before launch. The illustration also shows some sky areas covered by the surveys: {\it Kepler} (red), {\it Kepler}-K2 (green), TESS (Continuous Viewing Zones-CVZ; yellow) and CoRoT (magenta). Courtesy of Valerio Nascimbeni (INAF-OAPD, Italy), on behalf of the PLATO Mission Consortium.}
        \label{fig:plato_fields}
\end{figure}

\begin{figure}
        \begin{center}
                \includegraphics[height=7.0cm, width=\hsize]{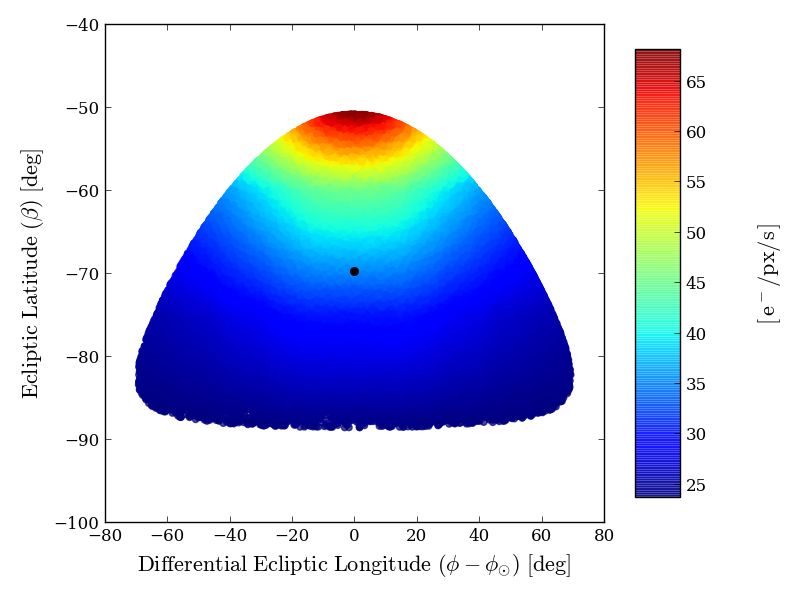}
        \end{center}
        \caption{Scatter plot of zodiacal light across IF in differential ecliptic coordinates centred on the Sun. Values are valid for an observer in L2 orbit having the Sun's ecliptic longitude $\phi_{\odot}$ aligned with $\phi_{LoS}=83.62^{\circ}$. Instrument vignetting is not included ($f_{vig}=0$).}
        \label{fig:stellar_field}
\end{figure}

With an nominal mission duration of four years, two observation scenarios are considered for PLATO. The first consists of two long-duration (2+2 years) observation phases (LOP) with distinct sky fields. The second consists of a single LOP of three years plus one step-and-stare operation phase (SOP) covering multiple fields lasting a few months each. Mission design constraints require the LOP fields to have absolute ecliptic latitude and declination above 63$^{\circ}$ and 40$^{\circ}$, respectively. Under such conditions, two LOP fields are actually envisaged: a southern PLATO field (SPF) centred at Galactic coordinates $l=253^{\circ}$ and $b=-30^{\circ}$ (towards the Pictor constellation) and a northern PLATO field (NPF) centred at $l=65^{\circ}$ and $b=30^{\circ}$ (towards the Lyra and Hercules constellations and also including the {\it Kepler} target field). An illustration containing the locations of both SPF and NPF is shown in Fig.~\ref{fig:plato_fields}, as well as the possible locations for the SOP fields, which will be defined two years before launch \cite[]{ESA2017_RBD}.
\begin{table}
        \caption[]{Coordinates of the input field line of sight ($\mathrm{IF_{LoS}}$) in different reference systems.}
        \label{table:los_coord}
        \centering
        \begin{tabular}{c c c}
                \hline
                \noalign{\smallskip}
                Reference system & Longitude & Latitude \\ [0.5ex]
                \hline 
                \noalign{\smallskip}
                Galactic & $\mathrm{l_{LoS}=253^{\circ}}$ & $\mathrm{b_{LoS}=-30^{\circ}}$  \\
                \noalign{\smallskip}
                Equatorial & $\mathrm{\alpha_{LoS}=86.80^{\circ}}$ & $\mathrm{\delta_{LoS}=-46.40^{\circ}}$  \\
                \noalign{\smallskip}
                Ecliptic & $\mathrm{\phi_{LoS}=83.62^{\circ}}$ & $\mathrm{\beta_{LoS}=-69.77^{\circ}}$ \\
                \hline                                    
        \end{tabular}
\end{table}
For the purposes of this paper, we adopted as input field (IF) the fraction of SPF that is equivalent to the area covered by a single PLATO camera centred at SPF centre. That represents roughly half of the SPF area in the sky and encompasses about 12.8 million stars listed in the {\it Gaia} DR2 catalogue with $G$ magnitude comprised between 2.45 and 21. Table~\ref{table:los_coord} presents, in different reference systems, the coordinates of IF centre, hereafter referred to as IF line of sight ($\mathrm{IF_{LoS}}$). The sky area covered by IF in ecliptic coordinates with zero point $\mathrm{\phi_{\odot}}$ in the Sun is illustrated in Fig.~\ref{fig:stellar_field}. We also present in the this figure the zodiacal light (see model description in Section~\ref{sec:sky_background}) perceived by an observer located in L2 orbit and pointing towards our coordinate zero point.

\subsection{Synthetic PLATO $P$ photometric passband}

\subsubsection{Definition and relationship with $V$ band}
The PLATO mission was designed based on stellar magnitudes specified in the visible band. Nevertheless, to avoid the inconvenience of having colour dependency when estimating stellar fluxes at detector level, from the visual magnitudes, it is more appropriate to work in a proper instrument photometric band. Therefore, we build in this paper a synthetic $P$ magnitude calibrated in the VEGAMAG system
\begin{equation}\label{eq:p_band}
P = -2.5 \log_{10}\left({
        \frac{\displaystyle\int_{\lambda_1}^{\lambda_2} f(\lambda) \, S(\lambda) \, \lambda \, d\lambda}
        {\displaystyle\int_{\lambda_1}^{\lambda_2} f_{\mathrm{Vega}}(\lambda) \, S(\lambda) \, \lambda \, d\lambda}
}\right) + P_{\mathrm{Vega}},
\end{equation}
where $f(\lambda)$ is the spectral flux of a given star, $f_{\mathrm{Vega}}(\lambda)$ is the spectral flux of the Vega A0V type star (see Fig.~\ref{fig:plato_throughput}), and $P_{\mathrm{Vega}}$ is its magnitude in the $P$ band, the latter assumed to be equal to $V_{\mathrm{Vega}} = 0.023$ mag \cite[]{Bohlin2007}. The $P$ band zero point is given by
\begin{equation}\label{eq:ZP_P}
zp = 2.5 \log_{10} \left( \left( h \, c \right)^{-1} \Theta \displaystyle\int_{\lambda_1}^{\lambda_2} f_{\mathrm{Vega}}(\lambda) \, S(\lambda) \, \lambda \, d\lambda \right) + P_{\mathrm{Vega}}.
\end{equation}
This constant (see Table~\ref{table:zero_ponits}) provides a straightforward way for switching between stellar flux and magnitudes using
\begin{equation}\label{eq:mag_to_flux_P}
P = -2.5 \log_{10} \left( \left( h \, c \right)^{-1} \Theta \displaystyle\int_{\lambda_1}^{\lambda_2} f(\lambda) \, S(\lambda) \, \lambda \, d\lambda \right) + zp.
\end{equation}
Thus, having the zero point $zp$ and the magnitude $P$ of a given star, its respective total flux $f_P$ (per camera and expressed in units of $\mathrm{e^- \, s^{-1}}$) can be estimated with
\begin{equation}\label{eq:flux_P}
f_P = 10^{-0.4 (P - zp)}.
\end{equation}
For switching between $P$ and $V$ magnitudes, we determine the $V-P$ relationship using the Johnson-Cousins $V$ filter (Fig.~\ref{fig:plato_throughput}) and modelling $f(\lambda)$ with synthetic stellar spectra extracted from the POLLUX database \cite[]{Palacios2010}. As for the calibration star Vega, we adopted the template $\mathrm{\texttt{alpha\_lyr\_stis\_008}}$ (Fig.~\ref{fig:plato_throughput}) from CALSPEC. The resulting $V-P$ samples are shown in Fig.~\ref{fig:color_indexes} as a function of the effective temperature $T_{\mathrm{eff}}$, the latter ranging from 4,000K to 15,000K in steps of 500K. The corresponding fitted polynomial is
\begin{equation}\label{eq:poly_fit_V_P_teff}
\begin{split}
V - P = -1.184 \times 10^{-12} (T_{\mathrm{eff}})^3 + 4.526 \times 10^{-8} (T_{\mathrm{eff}})^2 - \\
             5.805 \times 10^{-4} T_{\mathrm{eff}} + 2.449.
\end{split}
\end{equation}
Therefore, for a star with specified visual magnitude and $T_{\mathrm{eff}}$, we can determine its $P$ magnitude with Eq.~\ref{eq:poly_fit_V_P_teff} and then applied Eq.~\ref{eq:flux_P} to estimate the respective flux at detector level. Table~\ref{table:reference_star_flux} shows the predicted flux $f^{\mathrm{ref}}_{P}$ for a reference PLATO target, i.e. a 6,000K G0V spectral type star, as a function of its $V$ and $P$ magnitudes. The values include brightness attenuation due to vignetting for a source at $\alpha = 14^{\circ}$. In this scenario, a reference PLATO star with $V = 11$ has $P = 10.66$ and $f^{\mathrm{ref}}_{P} = 9.074 \, \mathrm{ke^- \, s^{-1}}$ per camera, or $\sim218 \, \mathrm{ke^- \, s^{-1}}$ when cumulating over 24 cameras.
\begin{table}
        \caption[]{Normal camera (N-CAM) predicted flux, $f^{\mathrm{ref}}_{P}$, for a reference 6,000K G0V star as a function of its $V$ and $P$ magnitudes. Values include vignetting for a source at $\alpha = 14^{\circ}$ (Table~\ref{table:plato_vignetting}) and are consistent with the current instrument design.}
        \label{table:reference_star_flux}
        \centering
        \begin{tabular}{c c c c}
                \hline
                \noalign{\smallskip}
                $V$   & $P$   & $f^{\mathrm{ref}}_{P}$ (per camera) & $f^{\mathrm{ref}}_{P}$ \textbf{(24 cameras)} \\ [0.5ex]
                [mag] & [mag] & $\mathrm{[10^3 \, e^-/s]}$                   & $\mathrm{[10^3 \, e^-/s]}$                            \\
                \hline 
                \noalign{\smallskip}
                8.0  & \textbf{7.66}  & 143.820 & \textbf{3,451.7} \\
                \noalign{\smallskip}
                8.5  & \textbf{8.16}  & 90.745 & \textbf{2,177.9} \\
                \noalign{\smallskip}
                9.0  & \textbf{8.66}  & 57.256 & \textbf{1,374.1} \\
                \noalign{\smallskip}
                9.5  & \textbf{9.16}  & 36.126 & \textbf{867.0} \\
                \noalign{\smallskip}
                10.0 & \textbf{9.66}  & 22.794 & \textbf{547.1} \\
                \noalign{\smallskip}
                10.5 & \textbf{10.16} & 14.382 & \textbf{345.2} \\
                \noalign{\smallskip}
                11.0 & \textbf{10.66} & 9.074 & \textbf{217.8} \\
                \noalign{\smallskip}
                11.5 & \textbf{11.16} & 5.726 & \textbf{137.4} \\
                \noalign{\smallskip}
                12.0 & \textbf{11.66} & 3.613 & \textbf{86.7} \\
                \noalign{\smallskip}
                12.5 & \textbf{12.16} & 2.279 & \textbf{54.7} \\
                \noalign{\smallskip}
                13.0 & \textbf{12.66} & 1.438 & \textbf{34.5} \\
                \hline                                    
        \end{tabular}
\end{table}

\subsubsection{Obtaining $P$ and $V$ from {\it Gaia} magnitudes}
\begin{figure}
        \begin{center}
                \includegraphics[height=17cm, width=8.5cm]{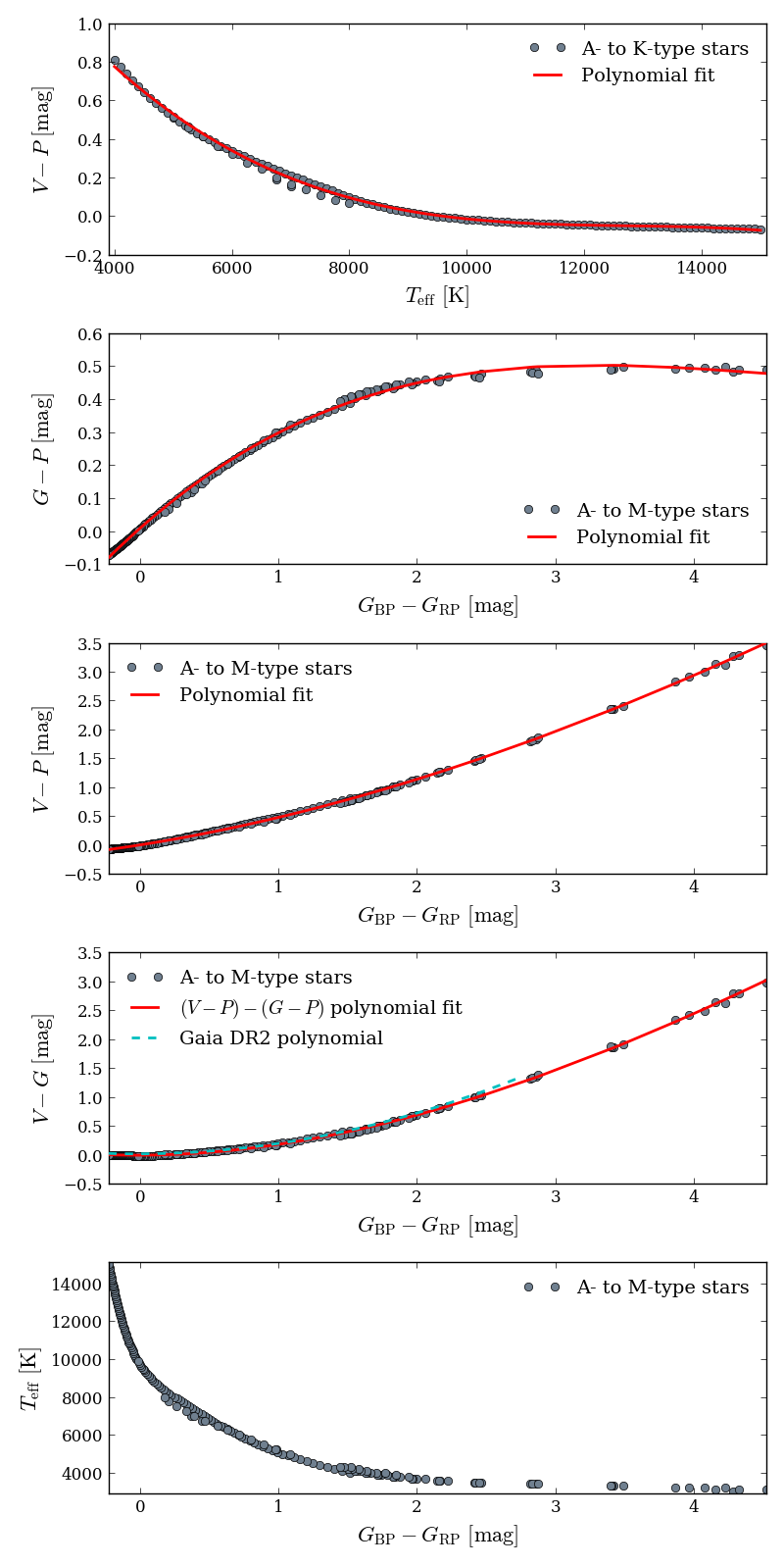}
        \end{center}
        \caption{Relationships between the photometric passbands $V$, $P$, and $G$, obtained by modelling $f(\lambda)$ with A- to M-type synthetic stellar spectra extracted from the POLLUX database. Red polynomials are derived from Eqs.~\ref{eq:poly_fit_V_P_teff}, ~\ref{eq:poly_fit_G_P_bp_rp}, and~\ref{eq:poly_fit_V_P_bp_rp}, and are applicable within the range $4,000 \leq T_{\mathrm{eff}} \leq 15,000$ (top frame) and $-0.227 \leq G_{BP}-G_{RP} \leq 4.524$. {\it Gaia} DR2 polynomial is based on Landolt stars observed with {\it Gaia} and is applicable within the range $-0.5 \leq G_{BP}-G_{RP} \leq 2.75$. It has a scatter of $4.6 \times 10^{-2}$ mag \cite[see][]{Busso2018}. The relationship between $T_{\mathrm{eff}}$ and our synthetic $G_{\mathrm{BP}} - G_{\mathrm{RP}}$ color index (bottom frame) is consistent with the color–temperature relations published in \cite{Andrae2018}.}     
        \label{fig:color_indexes}
\end{figure}

We also need to determine expressions for converting from the magnitude scales available in our adopted input catalogue ({\it Gaia} DR2) to our synthetic $V$ and $P$ magnitudes. {\it Gaia} collects data in three photometric systems: $G$, $G_{\mathrm{BP}}$, and $G_{\mathrm{RP}}$. As defined in \cite{Jordi2010}, all of these systems are calibrated in the VEGAMAG system, following therefore the same philosophy as Eqs.~\ref{eq:p_band},~\ref{eq:ZP_P}, and~\ref{eq:mag_to_flux_P}. To keep consistency with our previously adopted $V$ and $P$ bands, we applied the same Vega model to derive synthetic calibrations for the three {\it Gaia} bands. Consequently, we imposed to the latter the corresponding Vega magnitudes listed in \cite{Casagrande2018}. Table~\ref{table:zero_ponits} summarizes the obtained zero points for our synthetic $G$, $G_{BP}$, and $G_{RP}$ bands from this approach. They present satisfactorily low deviations with respect to the reference DR2 magnitudes published in \cite{Evans2018} and the later improved versions in \cite{Weiler2018}.
\begin{table}
        \caption[]{Zero points $zp$ of our synthetic $P$, $G$, $G_{BP}$, and $G_{RP}$ photometric passbands calibrated with Vega $\mathrm{\texttt{alpha\_lyr\_stis\_008}}$ model. Vega magnitudes for {\it Gaia} passbands are extracted from \cite{Casagrande2018}. Absolute deviations ($zp$ dev.) of $G$, $G_{BP}$, and $G_{RP}$ zero points are computed with respect to the reference DR2 magnitudes presented in \cite{Evans2018} (A) and the revised versions in \cite{Weiler2018} (B).}
        \label{table:zero_ponits}
        \centering
        \resizebox{\columnwidth}{!}{
        \begin{tabular}{c c c c c}
                \hline
                \noalign{\smallskip}
                Synthetic  & Vega   & $zp$   & $zp$ dev. (A) & $zp$ dev. (B) \\ [0.5ex]
                passband   & [mag]  & [mag]  & [mag] & [mag]\\
                \hline 
                \noalign{\smallskip}
                $P$ & 0.023 & 20.62 &  & \\
                \noalign{\smallskip}
                $G$ & 0.029 & 25.6879 & $4.6 \times 10^{-4}$ & $4.70 \times 10^{-2}$ \\
                \noalign{\smallskip}
                $G_{BP}$ & 0.039 & 25.3510 & $4.3 \times 10^{-4}$ & $1.10 \times 10^{-2}$ \\      
                \noalign{\smallskip}            
                $G_{RP}$ & 0.023 & 24.7450 & $1.69 \times 10^{-2}$ & $1.50 \times 10^{-2}$ \\
                \hline                                    
        \end{tabular}
        }
\end{table}
Then, to obtain both $P$ and $V$ magnitudes from the {\it Gaia} $G$ band, we determined $G-P$ and $V-P$ relationships by means of the $G_{\mathrm{BP}} - G_{\mathrm{RP}}$ colour index, resulting in the plots shown in Fig.~\ref{fig:color_indexes}. The corresponding fitted polynomials, within the range $-0.227 \leq G_{\mathrm{BP}} - G_{\mathrm{RP}} \leq 4.524$, are
\begin{equation}\label{eq:poly_fit_G_P_bp_rp}
\begin{split}
G - P = 0.00652 \, (G_{\mathrm{BP}} - G_{\mathrm{RP}})^3 - 0.08863 \, (G_{\mathrm{BP}} - G_{\mathrm{RP}})^2 + \\
                 0.37112 \, (G_{\mathrm{BP}} - G_{\mathrm{RP}}) + 0.00895;
\end{split}
\end{equation}
\begin{equation}\label{eq:poly_fit_V_P_bp_rp}
\begin{split}
V - P = -0.00292 \, (G_{\mathrm{BP}} - G_{\mathrm{RP}})^3 + 0.10027 \, (G_{\mathrm{BP}} - G_{\mathrm{RP}})^2 + \\
                 0.37919 \, (G_{\mathrm{BP}} - G_{\mathrm{RP}}) + 0.00267.
\end{split}
\end{equation}
Unlike Eq.~\ref{eq:poly_fit_V_P_teff}, the expressions in Eqs.~\ref{eq:poly_fit_G_P_bp_rp} and~\ref{eq:poly_fit_V_P_bp_rp} are described as a function of the $G_{\mathrm{BP}} - G_{\mathrm{RP}}$ colour index, rather than the $T_{\mathrm{eff}}$. The reason for that is the low availability of effective temperatures in DR2 (less than 10\% of the sources). In contrast, $G_{\mathrm{BP}}$ and $G_{\mathrm{RP}}$ magnitudes are simultaneously available for more than 80\% of the sources. To verify the consistency of our synthetic calibrations  derived from synthetic stellar spectra, we compared our $V-G = (V-P) - (G-P)$ relationship with the $V-G$ polynomial fit \cite[]{Busso2018} derived from Landolt\footnote{\url{https://www.eso.org/sci/observing/tools/standards/Landolt.html}.} standard stars (398 sources) observed with {\it Gaia}. As shown in Fig.~\ref{fig:color_indexes}, our synthetic $V-G$ curve exhibits satisfactory agreement with the $V-G$ polynomial fit obtained from the true {\it Gaia} observations. The maximum absolute error between both curves is $9.8 \times 10^{-2}$ mag at $G_{\mathrm{BP}} - G_{\mathrm{RP}} = 2.75$ mag. Hence, for the purposes of this paper, we consider that the polynomials of Eqs.~\ref{eq:poly_fit_G_P_bp_rp} and~\ref{eq:poly_fit_V_P_bp_rp} give sufficiently accurate estimates of $P$ and $V$ magnitudes from the $G$ magnitude of the DR2 catalogue.

\subsection{Identifying target and contaminant stars}\label{sec:stellar_sample}
\begin{figure}
        \begin{center}
                \includegraphics[height=11.0cm, width=\hsize]{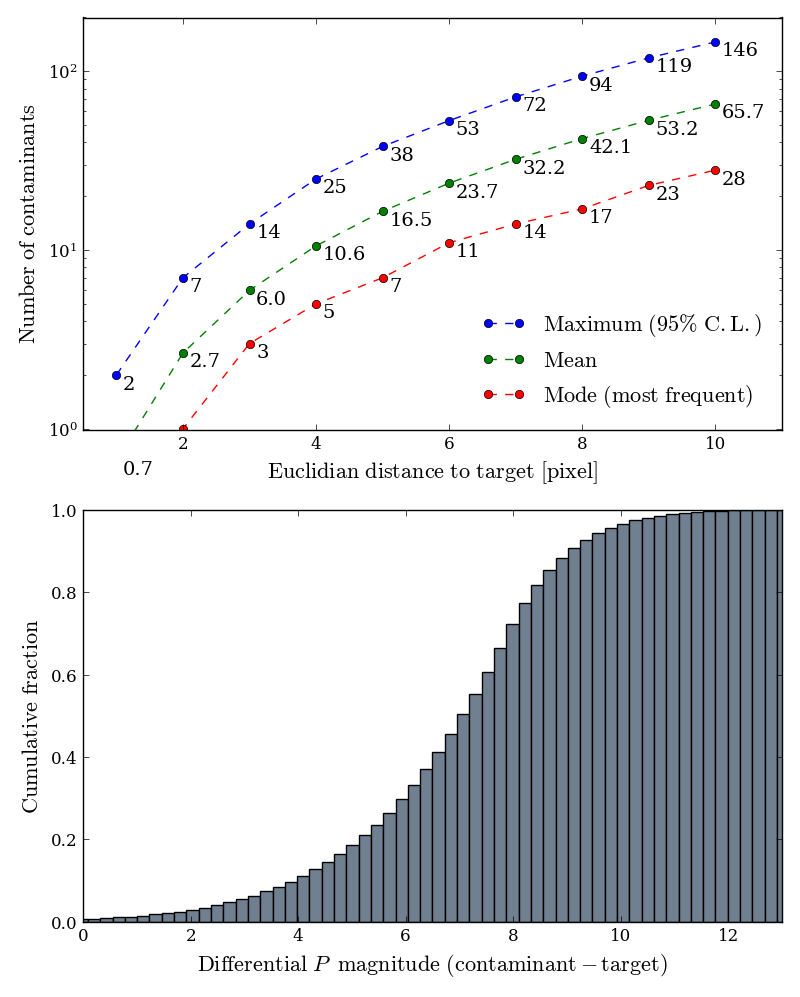}
                \caption{\textbf{Top}: Number of contaminants brighter than $P\sim21.1$ as a function of the Euclidean distance from the target stars ($7.66 \leq P \leq 12.66$). Maximum values have 95\% confidence level. \textbf{Bottom}: Cumulative fraction of the differential $P$ magnitude between contaminant and target stars, the former located at up to 10 pixels in distance from the latter.}
                \label{fig:target_contaminant_population}
        \end{center}
\end{figure}

We define in this section the ensemble of target and contaminant stars from the input catalogue that will be used to build input images for simulating aperture photometry. First, we determined the position of each star within IF at the focal plane array of one PLATO camera (as explained in Section~\ref{sec:input_field}, IF covers exactly the field of a single camera). Next, following the definition of the P5 sample, we assigned as targets those stars located within IF that have $0.57 \leq G_{BP}-G_{RP} \leq 1.84$ (F5 to late-K spectral types) and $P$ magnitude in the range $7.66 \leq P \leq 12.66$, the latter corresponding to $8.0 \leq V \leq 13.0$ for a reference PLATO target, i.e. a 6,000K G0V star. This accounts for about 127,000 sources. Target selection based on the $P$ band is more convenient than the $V$ band, as it allows us to overcome the colour dependency of the latter. In other words, this approach ensures that all targets assume flux values within a fixed range (that of Table~\ref{table:reference_star_flux}), regardless of their effective temperature. This is thus consistent with a target selection strategy driven by noise performance, magnitude, and spectral type. As for the contaminant stars, they correspond to all existing sources in the input catalogue located within 10 pixel radius around all targets. This accounts for about 8.3 million stars with $P$ magnitude comprised in the range $2.1 \lesssim P \lesssim 21.1$. It is important to mention that only sources satisfying $-0.227 \leq G_{BP}-G_{RP} \leq 4.524$ are included in the ensemble of contaminant stars to conform with the range of applicability of the polynomials described in Eqs.~\ref{eq:poly_fit_G_P_bp_rp} and~\ref{eq:poly_fit_V_P_bp_rp}. According to the above description, Fig.~\ref{fig:target_contaminant_population} presents some statistics (distances and differential magnitudes) on the distribution of contaminant stars relative to their corresponding targets.

A few considerations are necessary concerning the sources in {\it Gaia} DR2. \cite{Evans2018} reported some very likely saturation and imperfect background subtraction issues affecting sources with $G \lesssim 3.5$ and $G \gtrsim 17$, respectively. Since the central point in this work is to establish a relative performance comparison between different photometric aperture models -- particularly in scenarios of high stellar crowding -- we decided to not remove those sources from our working subset of stars. The inaccuracies resulting from the mentioned issues will ultimately be evenly propagated to all tested mask models, having therefore no potential to significantly impact the comparative basis analysis. 

\subsection{Setting up the input imagettes} \label{sec:input_imagettes}
\begin{figure}
        \begin{center}
                \includegraphics[height=5.0cm, width=5.5cm]{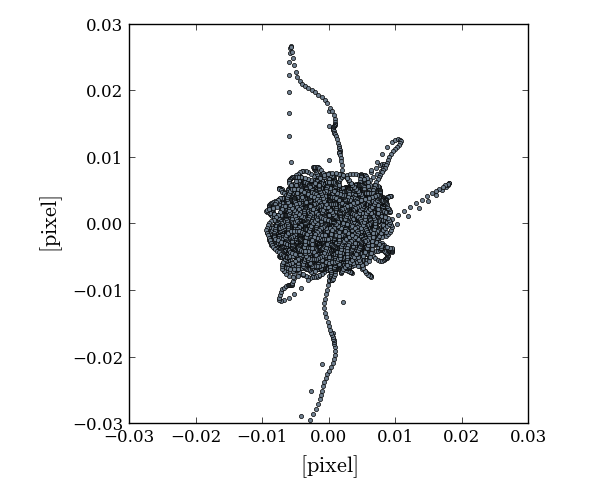}
                \caption{Illustration of a possible star position motion on the PLATO focal plane. We use as input a simulated time series of 1h duration sampled at 8Hz based on the current status of the pointing requirements. The Euclidian distance scatter is 2.25 mpixel rms with respect to the nominal (zero) position. Credit of the time series simulation: PLATO Industrial Core Team (OHB-System AG, TAS, RUAG Space).}
                \label{fig:pointing_time_series}
        \end{center}
\end{figure}

During science observations, in-flight photometry extraction will be performed independently for each target by integrating its flux over a set of selected pixels (aperture or mask). Such pixel collection is to be chosen from a $6\times6$ pixels square window called imagette, assigned uniquely to each target. An imagette is sufficiently large to encompass the near totality ($\sim99.99\%$) of the energy from the corresponding target. Characterizing the expected performance of mask-based photometry therefore requires building up such imagettes, which shall be composed of realistic stellar content (targets and respective contaminants). To do so, we applied the following procedure:
\begin{enumerate} \setlength{\itemsep}{4pt}
        \item Use the input catalogue and derived properties to obtain the magnitudes, fluxes, and locations of target and contaminant stars at intrapixel level. From that, we consider a random stellar subset composed of 50,000 targets (from the total of $\sim127$k potential targets within IF). These are neighboured by $\sim$3.25 million contaminant stars.
        \item Employ the PSFs presented in Section~\ref{sec:psf} as models of both instrument optical and detector responses to stellar flux.
        \item By convention, the pixels of an imagette are selected such that the centre of the resulting imagette is located at no more than an absolute Euclidean distance of 0.5 pixel from the target barycentre (see examples in Fig.~\ref{fig:psf_imagette}). This is done to maximize the amount of target energy falling within its imagette. 
        \item Translate each sample of the satellite pointing time series from Fig~\ref{fig:pointing_time_series} into a corresponding shifted imagette with respect to the nominal position (zero). These are used as input to compute the jitter noise in the photometry.  
\end{enumerate}
\begin{figure}
        \begin{center}
                \includegraphics[height=4.0cm, width=8.0cm]{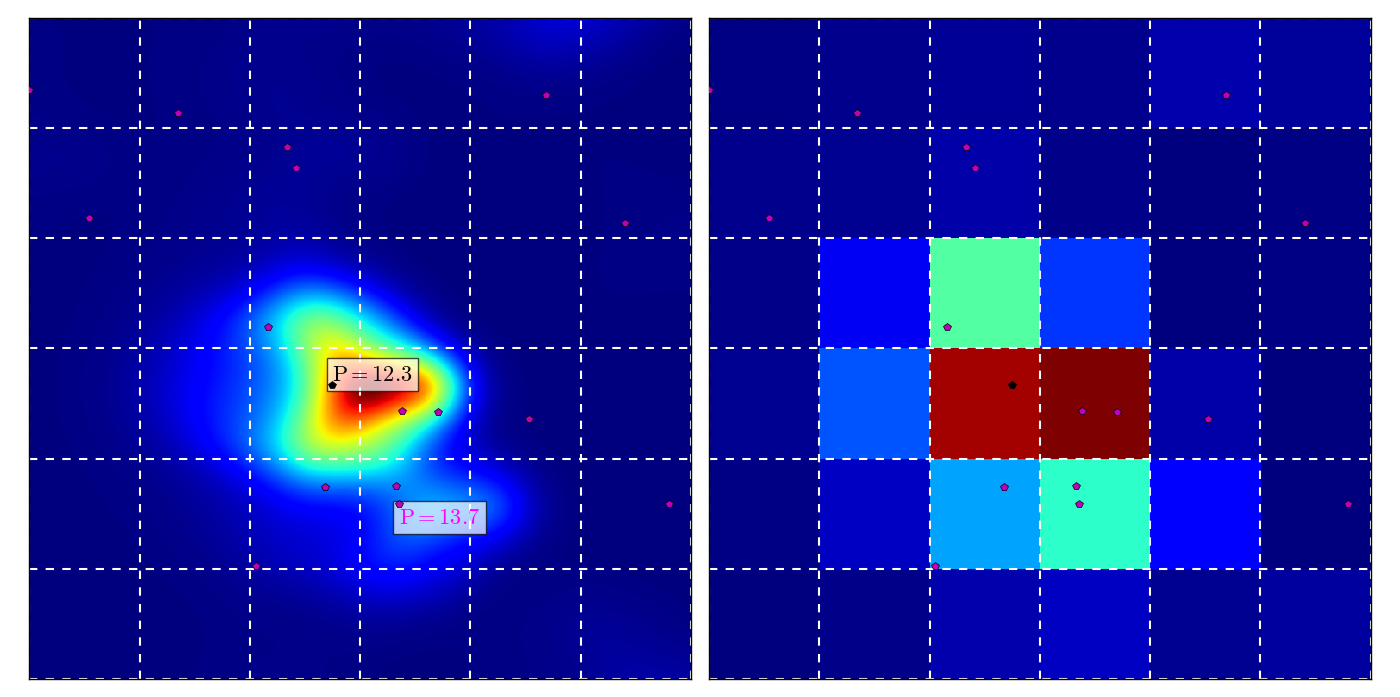}
                \caption{Example of input image. \textbf{Left}: High resolution PSF ($\alpha = 18^{\circ}$) for a target with $P = 12.3$ (barycentre designated by the black dot) surrounded by several contaminants (respective barycentres designated by the magenta dots). The brightest contaminant in the frame (tagged below the target) has $P=13.7$. All other contaminants are at least 2 mag fainter than the target. \textbf{Right}: Corresponding low resolution PSF (imagette) at pixel level. Zodiacal light is not shown in the scene. Dashed white lines represent pixel borders.}
                \label{fig:input_image_example}
        \end{center}
\end{figure}
Following this process, an input image (reference frame) like that illustrated in Fig.~\ref{fig:input_image_example} was generated for each target (including respective contaminants). Shifted images to account for satellite motion are therefore computed target by target with respect to their respective reference frames. To increase simulation speed, the jitter time series was down-sampled by a factor of 10, resulting in a 0.8 Hz signal keeping the same statistical properties (mean, variance, and spectral energy distribution) as the original signal. Based on that, a total of 2,880 shifted images (1h duration signal) were produced per reference image, resulting in a total of $50,000 \times (2880 + 1) \sim 144 \times 10^6$ synthetic imagettes. These are used as input for a detailed and realistic characterization of the performance expected from aperture photometry.

\section{Aperture photometry} \label{sec:aperture_photometry}
In order to find the optimal aperture model for extracting photometry from PLATO P5 targets, we applied the following steps:

i. Formalize an expression for the NSR.

ii. Define an expression for estimating the fractional flux from contaminant stars entering the apertures. This parameter is referred to as stellar pollution ratio (SPR).

iii. Build different aperture models based on NSR and width.

iv. Apply each mask model to the input images generated as described in Section~\ref{sec:input_imagettes}.

v. Compute, for each mask model, the number of target stars for which an Earth-like planet orbiting it would be detected, i.e, trigger a TCE. This parameter is referred to as  $N^{good}_\mathrm{TCE}$.

vi. Compute, for each mask model, the number of contaminant stars that are likely to produce, whenever they are eclipsed, background false positives. This parameter is referred to as $N^{bad}_\mathrm{TCE}$.

The above steps are detailed throughout the next sections in this chapter. Next, the performance of the different mask models are compared and commented on detail in Section~\ref{sec:results}.

\subsection{Noise-to-signal ratio} \label{sec:nsr}
The  NSR is the principal performance indicator for evaluating the exploitability of photometry signals. For PLATO stellar light curves derived from aperture photometry applied to imagettes, we used the following metric to compute the per cadence NSR ($\mathrm{NSR_{*}}$; see parameters description in Table~\ref{table:nsr_parameters}):
\begin{equation}\label{eq:nsr_cadence}
\mathrm{NSR_{*}} = 
\frac{
        \sqrt{
                \sum \limits_{n=1}^{36} \left( \sigma^2_{F_{T_{n}}} + \sum \limits_{k=1}^{N_C} \sigma^2_{F_{C_{n,k}}} + \sigma^2_{B_{n}} + \sigma^2_{D_{n}} + \sigma^2_{Q_{n}} \right) \, w^2_{n}
        }
}{
        \sum\limits_{n=1}^{36} F_{T_{n}} \, w_{n}
}.
\end{equation}
\begin{table}
        \caption[]{Description of the parameters of Eq.~\ref{eq:nsr_cadence}.}
        \label{table:nsr_parameters}
        \centering
                \begin{tabular}{c c c c}
                        \hline
                        \noalign{\smallskip}
                        Description & Symbol \\ [0.5ex]
                        \hline 
                        \noalign{\smallskip\smallskip}
                        Photon noise from the target star & $\sigma^2_{F_{T}}$ \\
                        \noalign{\smallskip\smallskip}
                        Photon noise from a contaminant star & $\sigma^2_{F_{C}}$ \\
                        \noalign{\smallskip\smallskip}
                        Background noise from the zodiacal light & $\sigma^2_{B}$ \\
                        \noalign{\smallskip\smallskip}
                        Overall detector noise & $\sigma^2_{D}$ \\
                        (including readout, smearing, and dark current) & \\
                        \noalign{\smallskip\smallskip}
                        Quantization noise & $\sigma^2_{Q}$ \\
                        \noalign{\smallskip\smallskip}
                        Average flux from the target star & $F_{T}$ \\
                        \noalign{\smallskip\smallskip}
                        Average flux from a contaminant star & $F_{C}$ \\
                        \noalign{\smallskip\smallskip}
                        Mask weight in the interval $[0,1]$ & $w$ \\
                        \noalign{\smallskip\smallskip}
                        imagette pixel index = $\{1,2,3,\dots,36\}$ & $n$ \\
                        \noalign{\smallskip\smallskip}
                        Contaminant star index = $\{1,2,3,\dots,N_C\}$ & $k$ \\
                        \noalign{\smallskip\smallskip}
                        Number of contaminant stars within & $N_C$ \\
                        10 pixel radius around the target & \\
                        \noalign{\smallskip}
                        \hline                                    
                \end{tabular}
\end{table}

A per cadence light curve sample corresponds to the integrated mask flux over one exposure interval of the detectors, which corresponds to 21 seconds (Table~\ref{table:instrument_data}) for PLATO N-CAM. In the context of PLATO, NSR scales over multiple independent samples and measurements,
\begin{equation} \label{eq:nsr}
\mathrm{NSR} = \frac{10^6}{12 \sqrt{t_{d} \, N_T}} \, \mathrm{NSR_{*}},
\end{equation}
where $t_d$ is the observation duration in hours and $N_T$ is the number of telescopes observing the star. The constant in the denominator of the above expression stands for the square root of the number of samples in one hour, i.e. $\sqrt{\mathrm{3600s/25s}} = 12$, based on the 25 seconds cadence (Table~\ref{table:instrument_data}) of the PLATO N-CAMs. For a signal with duration of one hour we use (expressed in units of $\mathrm{ppm \, hr^{1/2}}$)
\begin{equation} \label{eq:nsr_1h}
\mathrm{NSR_{1h}} = \frac{10^6}{12 \sqrt{N_T}} \, \mathrm{NSR_{*}}.
\end{equation}
We note that flux noise induced by satellite jitter is not included in Eq.~\ref{eq:nsr} at this stage. To do so would be a fairly complicated task because jitter contribution depends on the final shape of the aperture \cite[see][]{Fialho2007}. Later in this paper we explain how to include jitter noise in the NSR expressions, subsequent to the determination of the apertures.

\subsection{Stellar pollution ratio} \label{sec:spr}
We present herein the SPR. This factor permits us to quantify the average fractional contaminant flux from background stars captured by an aperture. We let $F_{C,k}$ be the photometric flux contribution from a single contaminant star $k$ and $F_{tot}$ the total flux. We have
\begin{eqnarray}
F_{C,k} = \sum\limits_{n=1}^{36} F_{C_{n,k}} \, w_{n}, \\
F_{tot} = \sum\limits_{n=1}^{36} \left( F_{T_{n}} + B_n + \sum \limits_{k=1}^{N_C} F_{C_{n,k}} \right) \, w_{n},
\end{eqnarray}
where $B_n$ is the average background flux at pixel $n$ from the zodiacal light. We denote $\mathrm{SPR}_k$ as the fractional flux from the contaminant star $k$ with respect to the total photometric flux (target plus contaminants and zodiacal light), i.e.
\begin{equation}\label{eq:SPR_k}
\mathrm{SPR}_{k} = \frac{F_{C,k}}{F_{tot}}.
\end{equation}
Accordingly, the fractional flux from all contaminant stars is
\begin{equation}\label{eq:SPR_total}
\mathrm{SPR}_{tot} = \sum \limits_{k=1}^{N_C} \mathrm{SPR}_{k}.
\end{equation}
We note that $\mathrm{SPR}_{tot}$ is complementary to the crowding metric $r$ defined in \cite{Batalha2010}, i.e. $\mathrm{SPR}_{tot} = 1 - r$.

\subsection{Detectability of planet transits}
When a planet eclipses its host star, it produces a maximum transit depth $\mathrm{\delta}_p$ which is, at first order approximation, equal to the square of the ratio between the planet radius and the star radius
\begin{equation} \label{eq:planet_transit_depth}
\delta_p = \left( R_p/R_{\star} \right)^2.
\end{equation}
In practice, $\mathrm{\delta}_p$ is always diluted by the contaminant flux from surrounding stars and background light, such that the observed transit depth $\delta_{obs}$ is a fraction of the original transit depth $\delta_p$
\begin{equation} \label{eq:observed_transit_depth_1}
\delta_{obs} = (1-\mathrm{SPR}_{tot}) \, \delta_p.
\end{equation}
Traditionally, a planet detection is not considered scientifically exploitable unless it has been observed at least three times. Furthermore, observed transits must reach a certain level of statistical significance, $\eta$, of the total noise, $\sigma$. In this paper, we adopted the threshold\footnote{This criterion was established to ensure that no more than one false positive due to random statistical fluctuations occurs over the course of the {\it Kepler} mission \cite[]{Jenkins2010}.} of $7.1\sigma$ ($\eta_{min} = 7.1$) as a minimum condition for characterizing a TCE with three transits. It yields
\begin{equation} \label{eq:observed_transit_depth_2}
\delta_{obs} \geq \eta_{min} \, \sigma = 7.1 \sigma.
\end{equation}
The total noise $\sigma$ scales with the signal (transit) duration $t_d$ and with the number of transit events $n_{tr}$, resulting
\begin{equation} \label{eq:sigma}
\sigma = \mathrm{NSR_{1h}} / \sqrt{t_d \, n_{tr}}.
\end{equation}
By combining the above expressions we can determine the range of detectable planet radius \cite[cf.][]{Batalha2010}
\begin{equation} \label{eq:Rp}
R_p \geq R_\star \sqrt{ \dfrac{\eta}{(1-\mathrm{SPR}_{tot})} \, \dfrac{\mathrm{NSR_{1h}}}{\sqrt{t_d \, n_{tr}}} }.
\end{equation}
Earth-like planets located at about 1au from Sun-like stars have $\delta_p \sim 84$ ppm and $t_d \sim 13$ hours. Consequently, it is required that $\mathrm{NSR_{1h}} \lesssim 74 \mathrm{ppm \, hr^{1/2}}$ for that type of planet to be detected at $\eta=\eta_{min} = 7.1$, $n_{tr}=3$ and $\mathrm{SPR}_{tot}=0$. From Eq.~\ref{eq:Rp}, we can obtain the statistical significance $\eta$ at which a planet can be detected
\begin{equation} \label{eq:eta}
\eta = \delta_p \sqrt{t_d \, n_{tr}} \left( 1 - \mathrm{SPR}_{tot} \right) / \mathrm{NSR_{1h}}.
\end{equation}
Therefore, an aperture model providing the highest number of targets stars with $\eta \geq \eta_{min}$ (i.e. highest $N^{good}_\mathrm{TCE}$), for $n_{tr} \geq 3$, is that being more likely in a statistical sense to detect true planet transits.

\subsection{Sensitivity to background false transits}
In this section, we derive a metric to evaluate the sensitivity of an aperture in detecting false planet transits originating from astrophysical eclipses of contaminant stars. Such events may occur, in particular, when the contaminant star in question is part of an  EB system and is sufficiently bright and sufficiently close to a target star. False planet transits caused by grazing EBs are thus not addressed herein.

When a given contaminant star $k$ is eventually eclipsed, we observe in the raw photometry a within aperture fractional flux decrease $\Delta F^{raw}_{C,k}$ and a corresponding within aperture fractional magnitude increase $\mathrm{\Delta m}^{raw}_{C,k}$, such that
\begin{equation} \label{eq:delta_m_eclipse}
\mathrm{\Delta m}^{raw}_{C,k} = -2.5 \log_{10} \Bigg( \frac{F^{raw}_{C,k} - \Delta F^{raw}_{C,k}}{F^{raw}_{C,k}} \Bigg).
\end{equation}
By denoting $\mathrm{\Delta m}^{raw}_{C,k}$ as the background transit depth $\delta_{back, k}$ in mag units and $\Delta F^{raw}_{C,k} / F_{tot}$ as the resulting observed transit depth $\delta_{obs, k}$ in the raw light curve, relative to the contaminant star $k$, we obtain
\begin{equation} \label{eq:delta_obs}
\delta_{obs, k} = \mathrm{SPR}^{raw}_{k} \left( 1 - {10}^{-0.4 \delta_{back, k}} \right), \\
\end{equation}
with
\begin{equation} \label{eq:spr_raw}
\mathrm{SPR}^{raw}_{k} = \frac{ F^{raw}_{C,k} }{ F_{tot} }.
\end{equation}
This expression shows that the background transit depth $\delta_{back, k}$ affects the light curve as an observed transit depth $\delta_{obs, k}$, which is proportional to $\mathrm{SPR}^{raw}_{k}$, i.e. the SPR of the contaminant star $k$ in the raw photometry. Because $\delta_{obs, k}$ is the result of a false planet transit, we want it to be sufficiently small to prevent it triggering a TCE, i.e.
\begin{equation} \label{eq:delta_obs_max}
\delta_{obs, k} < \eta_{min} \, \sigma.
\end{equation}
Although the above statement holds if, and only if, the  $\mathrm{SPR}_{k}$ is below a certain level for given $\delta_{back, k}$, $\eta$, $t_d$, and $n_{tr}$. We denote such a threshold as the critical SPR ($\mathrm{SPR}^{crit}_{k}$) of the contaminant star $k$. It can be determined with
\begin{equation} \label{eq:spr_critical}
\mathrm{SPR}^{crit}_{k} = \frac{\eta}{\left( 1-{10}^{-0.4 \delta_{back, k}} \right)} \, \frac{\mathrm{NSR_{1h}}}{\sqrt{t_d \, n_{tr}}}.
\end{equation}
Therefore, an aperture model providing the lowest number of contaminant stars for which $\mathrm{SPR}^{raw}_{k} \geq \mathrm{SPR}^{crit}_{k}$ (i.e. lowest $N^{bad}_\mathrm{TCE}$), for $\eta \geq \eta_{min} = 7.1$ and $n_{tr} \geq 3$, is that more likely in a statistical sense to naturally reject false planet transits caused by background eclipsing objects.

\subsection{Background flux correction} \label{sec:background_corection}
Background correction refers to subtracting, from the raw photometry, flux contributions from contaminant sources and scattered stray light (e.g. zodiacal and Galactic lights). The spatial distribution of background light is commonly describe using polynomial models, whose coefficients are determined based on flux measurements taken at strategically selected pixels \cite[see e.g.][]{Drummond2008, Twicken2010}. For PLATO, the strategy for background correction is not yet characterized at the present date, thus no accurate information on this subject is available for inclusion in our study. Notwithstanding, we investigate in this section what would be the impact of an ideally perfect background correction on the science metrics $N^{good}_\mathrm{TCE}$ and $N^{bad}_\mathrm{TCE}$. We assume therefore a hypothetical scenario in which $B_n = F_{C,k} = \mathrm{SPR}_{k} = \mathrm{SPR}_{tot} = 0$.

In this case, the observed depth of a legitimate planet transit simply converges to its true depth, i.e. $\delta_{obs} = \delta_p$ (the transit dilution is completely cancelled). In parallel, the parameter $\eta$ (Eq.~\ref{eq:eta}) increases, meaning that the apertures become more sensitive to detect true planet transits, which ultimately implies an increase in $N^{good}_\mathrm{TCE}$ as well.
        
Analysing the impact on $N^{bad}_\mathrm{TCE}$ is not as straightforward as it is for $N^{good}_\mathrm{TCE}$. First, we denote hereafter $F^{corr}_{tot}$ as the total photometric flux resulted after the background correction, which only contains signal from the target
\begin{equation}\label{eq:flux_photometry_corrected}
F^{corr}_{tot} = \sum\limits_{n=1}^{36} F_{T_{n}} \, w_{n}.
\end{equation}

Next, we denote $\Delta F^{raw}_{C,k} / F^{corr}_{tot}$ as the resulting observed transit depth $\delta^{corr}_{obs, k}$, after background correction, caused by an eclipse of the contaminant star $k$. This leads us, using Eq.~\ref{eq:delta_m_eclipse}, to an expression for $\delta^{corr}_{obs, k}$ which is similar to that of Eq.~\ref{eq:delta_obs}, except that the term ($F^{raw}_{C,k} / F^{corr}_{tot}$) appears in place of $\mathrm{SPR}^{raw}_{k}$, resulting in
\begin{equation} \label{eq:delta_obs_corr_a}
\delta^{corr}_{obs, k} = \left( F^{raw}_{C,k} / F^{corr}_{tot} \right) \left( 1 - {10}^{-0.4 \delta_{back, k}} \right).
\end{equation}
The above identity shows that removing the background flux from the photometry does not suppress the false transit caused by a background EB. Indeed, although the average flux from the eclipsing contaminant star goes to zero ($F_{C,k}=0$) in the corrected photometry, the transit depth $\delta^{corr}_{obs, k}$ depends on the intrinsic (raw) contaminant flux $F^{raw}_{C,k}$ that is present in the scene, which is thus independent of any further processing applied in the photometry. Besides, this result is consistent with the fact that the background correction only removes the nominal (out-of-transit) average flux of the contaminant source from the photometry, therefore becoming no longer effective if such signal changes after the correction (e.g. owing to an eclipse, i.e. when $\delta_{back, k}\neq0$).

For convenience, we define herein the apparent SPR ($\mathrm{SPR}^{app}_{k}$), which is manifested during the eclipse of a contaminant star $k$ in a light curve with flux fully corrected for the background
\begin{equation} \label{eq:spr_app}
\mathrm{SPR}^{app}_{k} = \frac{ F^{raw}_{C,k} }{ F^{corr}_{tot} }.
\end{equation}
This yields
\begin{equation} \label{eq:delta_obs_corr_b}
\delta^{corr}_{obs, k} = \mathrm{SPR}^{app}_{k} \left( 1 - {10}^{-0.4 \delta_{back, k}} \right).
\end{equation}
Comparing Eqs.~\ref{eq:delta_obs} and~\ref{eq:delta_obs_corr_b}, we note that $\delta^{corr}_{obs, k}$ is greater than $\delta_{obs, k}$, since $\mathrm{SPR}^{app}_{k} > \mathrm{SPR}^{raw}_{k}$. This means that the apertures become more sensitive to detect false planet transits from background eclipsing objects when the corresponding photometry is corrected for the average background flux. This happens because the background correction reduces the dilution of such transits.
From all the above considerations, it is possible to state therefore that the background correction is expected to increase both $N^{good}_\mathrm{TCE}$ and $N^{bad}_\mathrm{TCE}$ metrics.

\subsection{Aperture models} \label{sec:aperture_models}
From a purely scientific point of view on planet detection, an ideal aperture is that which is fully sensitive to all true, and fully insensitive to all false, planet transits. However, apertures cannot perfectly disentangle the flux of targets from that of their contaminant sources, so the ideal mask is physically impossible to achieve. Indeed, Eqs.~\ref{eq:eta} and~\ref{eq:spr_critical} show us that maximizing the yield of true planet transits and minimizing the occurrences of false planet transits are conflicting objectives: the former requires minimizing NSR and the latter maximizing it. Therefore, the concept of optimal aperture, in the context of this work, is defined as offering the best compromise regarding these two facets, even if the priority is of course to maximize the probability of finding true planet transits. With that in mind, we present in this section three mask models, each having a different shape and thus supplying distinct performance in terms of NSR and SPR. This gives us elements to check whether a solution giving overall best NSR also has satisfactory performance in terms of SPR and vice versa.

\subsubsection{Gradient mask} \label{sec:gradient_mask}
As NSR is the main performance parameter to be evaluated, a logical mask model to experiment with is that having weights $w_{n}$ providing the best $\mathrm{NSR}_{*}$ for each
target. Since the masks have by definition the same dimension of the imagettes, i.e. modest $6\times6$ pixels, it would be suitable to compute the collection of pixels providing minimum NSR by exhaustive search, i.e. by simple trials of several $w_{n}$ combinations, keeping that with lowest $\mathrm{NSR}_{*}$. Naturally, that kind of approach is far from efficient, especially considering that this procedure must be executed for tens of thousands of target stars. To avoid this inconvenience, we developed a direct method for calculating $w_{n}$ giving the best NSR. To determine such a mask, we rely on the fact that $\mathrm{NSR}_{*}$, at its minimum, should have a gradient identically equal to zero ($\mathrm{\nabla \mathrm{NSR}_{*} = 0}$) with respect to the weights. From this, we obtain 36 non-linear equations of the form
\begin{equation}\label{eq:nsr_grad_equations}
w_{n} \, \sigma^2_{n} \sum_{i=1}^{36} w_{i} \, F_{T_{i}} = F_{T_{n}} \sum_{i=1}^{36} w^2_{i} \, \sigma^2_{i},
\end{equation}
where $i$ is the imagette pixel index $=\{1,2,3,\dots,36\}$. One simple solution beyond the trivial result for $w_{n}$ satisfying the above equality can be calculated directly with
\begin{equation}\label{eq:nsr_grad_algo}
w_n = \dfrac{ F_{T_n} }{ \sigma^2_n }.
\end{equation}
Conventionally, all $w_{n}$ are then normalized by $\max_{} [w_n]$ to satisfy $0 \leq w_n \leq 1$, so that each weight $w_n$ directly represents the fraction of the imagette flux being caught by the aperture at the corresponding pixel $n$. For illustration, Fig.~\ref{fig:masks_examples}a shows the resulting gradient mask for the input image example of Fig.~\ref{fig:input_image_example}.

In order to simplify our terminology, the masks $w_n$ obtained from Eq.~\ref{eq:nsr_grad_algo} are hereafter referred to as gradient masks based on the fact that they are determined from the mathematical gradient of $\mathrm{NSR}_{*}$ expression. Each time they are mentioned however we should keep in mind that they correspond to the masks providing the global minimum NSR from all the possible combinations of mask weights $w_n$ in Eq.~\ref{eq:nsr_cadence}.

\subsubsection{Gaussian mask} \label{sec:gaussian_mask}
Having examined the shape of gradient masks applied to several stars, we noticed that they look very similar to a bell shaped curve. Therefore, we decided to test Gaussian-like masks to verify whether they could provide near-best NSRs when compared to gradient masks. Depending on the performance difference, the advantage of having an analytical mask that requires fewer parameters to be computed could justify its choice over the gradient mask. On these terms, we calculate the weights $w_n$ of a Gaussian mask using the conventional symmetric Gaussian function expression
\begin{equation}\label{eq:gaussian_masl}
        w_{x,y} = \exp{ \left( -\frac{ \left( x - x_{\star} \right)^2 + \left( y - y_{\star} \right)^2}{2 \sigma^2_{w}} \right) },
\end{equation}
where
\[
\begin{array}{lp{0.8\linewidth}}
(x, y)                     & are Cartesian coordinates of the imagette pixels with shape $6 \times 6$;\\
\noalign{\smallskip \smallskip}
(x_{\star}, y_{\star})     & are the coordinates of the target barycentre within the imagette;\\
\noalign{\smallskip \smallskip}
\sigma_{w}                 & is the mask width in pixels on both $x$ and $y$ dimensions;\\
\noalign{\smallskip \smallskip}
w_{x, y}                   & is the mask weight in the interval $[0,1]$ at $(x, y)$.\\
\end{array}
\]
As the imagette dimension is fixed and the target position within it is well known thanks to the input catalogue, choosing a Gaussian mask for a given target reduces to finding a proper width. For that, we simply iterate over different values of $\sigma_{w}$  and keep that giving the lowest $\mathrm{NSR_{*}}$, as shown in Fig.~\ref{fig:nsr_curves}.  For illustration, Fig.~\ref{fig:masks_examples}b shows the resulting best NSR Gaussian mask for the input image example of Fig.~\ref{fig:input_image_example}.

\subsubsection{Binary mask} \label{sec:binary_mask}
\begin{figure}
        \begin{center}
                \includegraphics[height=4.4cm, width=\hsize]{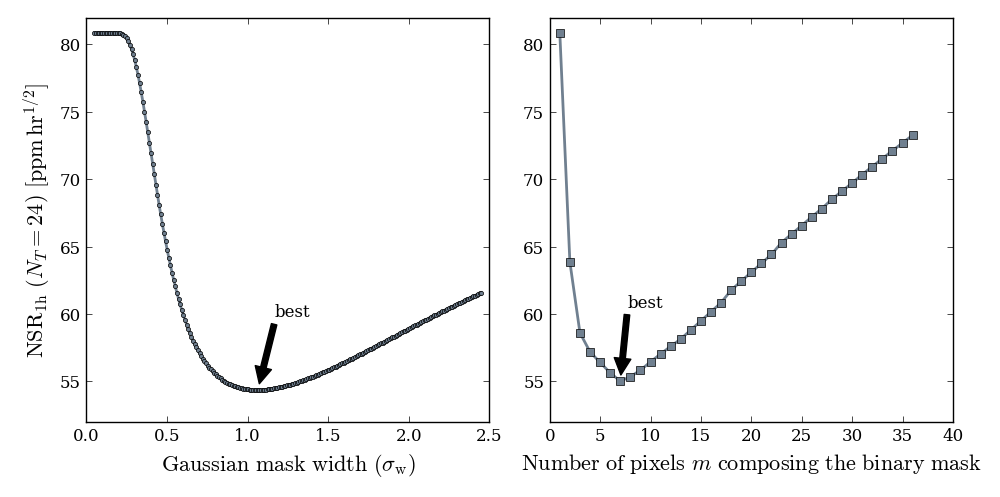}
                \caption{Example of NSR evolution curve as a function of the increasing aperture size for a target star with $P=11$. \textbf{Left}: Gaussian mask. \textbf{Right}: Binary mask.}
                \label{fig:nsr_curves}
        \end{center}
\end{figure}

\begin{figure}
        \begin{center}
                \includegraphics[height=3.0cm, width=\hsize]{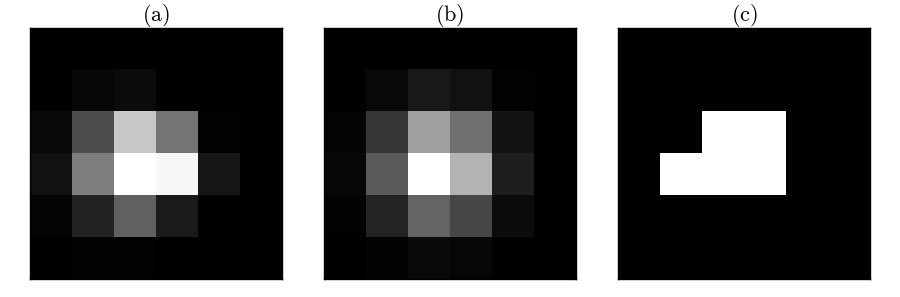}
                \caption{Aperture shapes computed as described in Section~\ref{sec:aperture_models}, for the input image example of Fig.~\ref{fig:input_image_example}. \textbf{Left (a)}: Gradient mask. \textbf{Centre (b)}: Gaussian mask. \textbf{Right (c)}: Binary mask.}
                \label{fig:masks_examples}
        \end{center}
\end{figure}

Binary masks are non-weighted apertures, meaning that the photometry is extracted by fully integrating pixel fluxes within the mask domain and discarding those which are outside it. This type of aperture was extensively employed to produce light curves of CoRoT and {\it Kepler} targets, which are thus well known for delivering satisfactory performance. In the context of PLATO, we applied the following routine to compute a binary mask for each target imagette.
\begin{enumerate}
        \item Arrange all pixels $n$ from the target imagette in increasing order of  $\mathrm{NSR}_n$
        
        \begin{equation}\label{eq:nsr_n}
        \mathrm{NSR}_n = 
        \frac{
                \sqrt{
                         \sigma^2_{F_{T_{n}}} + \sum \limits_{k=1}^{N_C} \sigma^2_{F_{C_{n,k}}} + \sigma^2_{B_{n}} + \sigma^2_{D_{n}} + \sigma^2_{Q_{n}}
                }
        }{
                F_{T_{n}}
        }.
        \end{equation}
        
        \item Compute the aggregate noise-to-signal $\mathrm{NSR}_{agg}(m)$, as a function of the increasing number of pixels $m=\{1,2,3,\dots,36\}$, stacking them to conform to the arrangement in the previous step and starting with the pixel owning the smallest $\mathrm{NSR}_n$
        \begin{equation}\label{eq:nsr_agg}
        \mathrm{NSR}_{agg}(m) = 
        \frac{
                \sqrt{
                        \sum \limits_{n=1}^{m} \left( \sigma^2_{F_{T_{n}}} + \sum \limits_{k=1}^{N_C} \sigma^2_{F_{C_{n,k}}} + \sigma^2_{B_{n}} + \sigma^2_{D_{n}} + \sigma^2_{Q_{n}} \right)
                }
        }{
                \sum\limits_{n=1}^{m} F_{T_{n}}
        }.
        \end{equation}
        \item Define as the aperture the collection of pixels $m$ providing minimum $\mathrm{NSR}_{agg}(m)$.
\end{enumerate}
As the binary mask gets larger following the above routine, the NSR typically evolves as illustrated in Fig.~\ref{fig:nsr_curves}. Accordingly, the resulting best NSR binary mask for the input image example of Fig.~\ref{fig:input_image_example} is shown in Fig.~\ref{fig:masks_examples}c.


\section{Performance assessment} \label{sec:results}
We present in this section the photometric performance of the three aperture models defined in Section~\ref{sec:aperture_models}. The results are presented in terms of NSR, SPR, number $N^{good}_\mathrm{TCE}$ of target stars with sufficiently low NSR permitting the detection of planets orbiting them, and number $N^{bad}_\mathrm{TCE}$ of contaminant stars with sufficiently high SPR to produce, should they be eclipsed, false positives. The results were obtained by applying each aperture model to all 50,000 input imagettes from Section~\ref{sec:input_imagettes}.

\subsection{Noise-to-signal ratio} \label{sec:nsr_results}
As already pointed out in Section~\ref{sec:nsr}, the per cadence  $\mathrm{NSR}_{*}$ from Eq.~\ref{eq:nsr} does not include photometric flux noise induced by spacecraft jitter because of its dependency on aperture weights. Once the apertures are computed however, we can include jitter noise in the photometry using the shifted imagettes described in Section~\ref{sec:input_imagettes}. We denote $\mathrm{NSR}^{jitter}_\mathrm{*}$ the per cadence NSR, which includes star motion due to satellite jitter, i.e.
\begin{equation}\label{eq:nsr_jitter}
\mathrm{NSR}^{jitter}_\mathrm{*} = \mathrm{NSR_{*}} \sqrt{ 1 + \left( \frac{\sigma^2_J}{\sigma^2_{*}} \right) },
\end{equation}
where $\sigma_{J}$ is the photometric jitter noise obtained from the shifted imagettes and $\sigma_{*}$ corresponds to the numerator of the expression in Eq.~\ref{eq:nsr_cadence}. The above expression considers stationary random noise for both photometric flux and satellite jitter. Table~\ref{table:jitter_factor} shows the impact of spacecraft jitter on the photometry under nominal and degraded scenarios of pointing performance. We verified that in nominal conditions the impact of jitter on the photometry is negligible, showing that including jitter in the calculation scheme of the apertures would not only represent a complicated procedure, but also a useless effort in that particular case.
\begin{table}
        \caption[]{Maximum noise-to-signal degradation at 95\% confidence level as a function of satellite jitter amplitude, computed from a sample of 10,000 targets. Four scenarios are considered: nominal (Fig.~\ref{fig:pointing_time_series}), three times (3$\times$) nominal, five times (5$\times$) nominal, and seven times (7$\times$) nominal jitter.}
        \label{table:jitter_factor}
        \centering
        \resizebox{\columnwidth}{!}{
                \begin{tabular}{c | c c c c}
                        \hline
                        \noalign{\smallskip}
                        Aperture model & Nominal & 3$\times$ Nominal & 5$\times$ Nominal & 7$\times$ Nominal \\ [0.5ex]
                        \hline 
                        \noalign{\smallskip}
                        Gradient & $0.31\%$ & $2.9\%$ & $8.1\%$ & $16.2\%$ \\
                        \noalign{\smallskip}
                        Gaussian & $0.41\%$ & $3.6\%$ & $10.0\%$ & $19.43\%$ \\
                        \noalign{\smallskip}
                        Binary   & $0.50\%$ & $4.7\%$ & $12.3\%$ & $23.2\%$ \\
                        \hline                                    
                \end{tabular}
        }
\end{table}

\begin{figure}
        \begin{center}
                \includegraphics[height=9.0cm, width=\hsize]{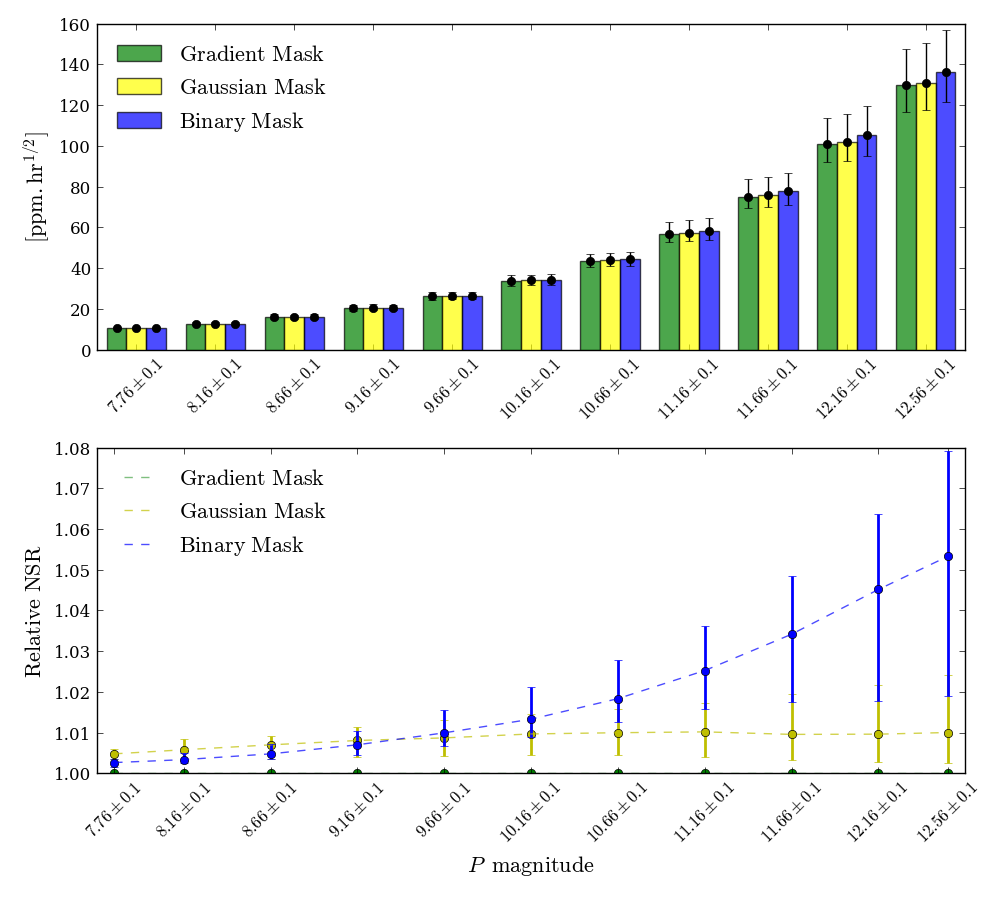}
                \caption{\textbf{Top}: Median values (black dots) of $\mathrm{NSR}^{jitter}_\mathrm{1h}$ ($N_T=24$) as a function of target $P$ magnitude and the applied mask model. \textbf{Bottom}: Relative $\mathrm{NSR}^{jitter}_\mathrm{1h}$, where the unit stands for the best NSR. In both plots, interval bars represent dispersions at 90\% confidence level.}
                \label{fig:nsr_results}
        \end{center}
\end{figure}

The performance parameter $\mathrm{NSR}^{jitter}_\mathrm{1h}$ was computed for our subset of input images assuming a nominal satellite jitter. The results are shown in Fig.~\ref{fig:nsr_results}. Overall, the three aperture models present comparable results for targets brighter than $P\sim 10.5$, with differences of less than 2\% on average. The Gaussian mask has consistent suboptimal NSR performance over the entire P5 magnitude range, that is only $\sim 1\%$ higher on average than the gradient mask. The binary mask has better performance on average than the Gaussian mask for targets brighter than $P\sim 9$, but its performance degrades rapidly with increasing magnitude. For the faintest P5 targets, the binary mask presents NSR values about 6\% higher on average and $\sim8\%$ higher in the worst scenarios with respect to the gradient mask. Therefore, looking exclusively in terms of NSR, weighted masks are clearly the best choice. 

\subsection{Stellar pollution ratio} \label{sec:spr_results}
\begin{figure}
        \begin{center}
                \includegraphics[height=5.2cm, width=\hsize]{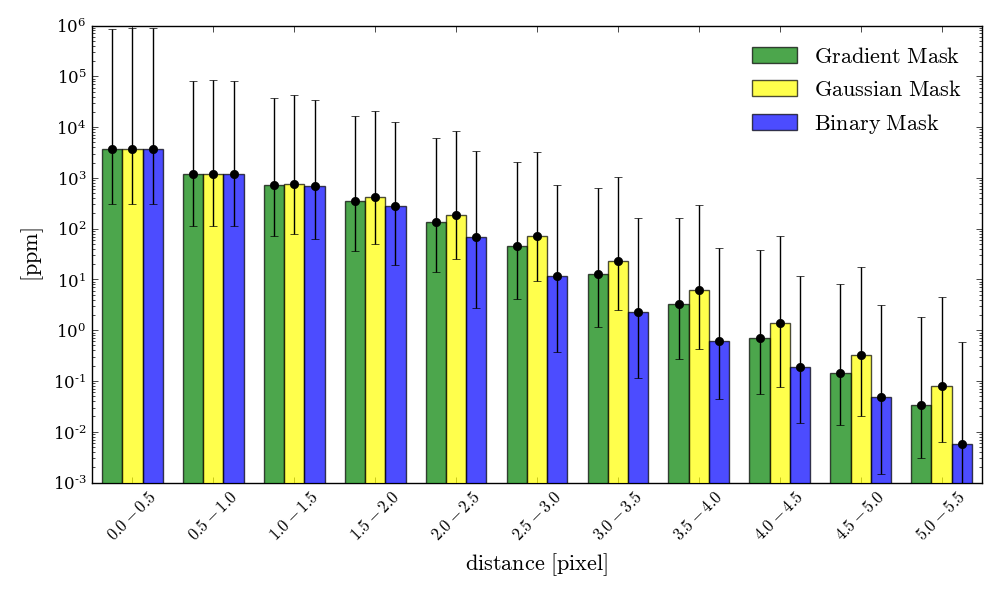}
                \caption{Median values (black dots) of $\mathrm{SPR}_{k}$ (Eq.~\ref{eq:SPR_k}) as a function of the distance in pixels between the contaminants sources and their respective targets, and the applied mask model. Interval bars represent dispersions at 90\% confidence level.}
                \label{fig:spr_k_results}
        \end{center}
\end{figure}

\begin{figure}
        \begin{center}
                \includegraphics[height=5.5cm, width=\hsize]{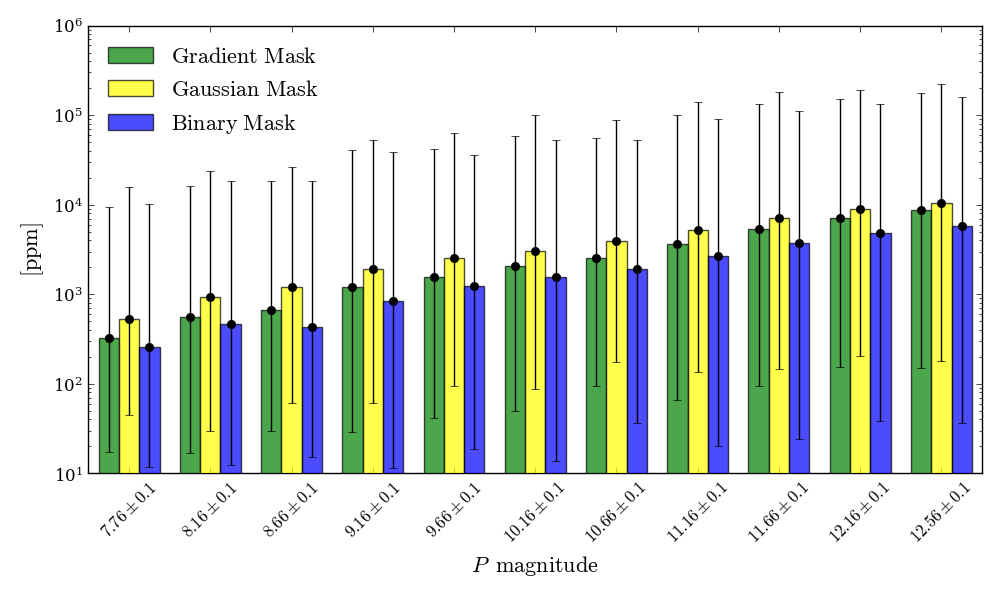}
                \caption{Median values (black dots) of $\mathrm{SPR}_{tot}$ (Eq.~\ref{eq:SPR_total}) as a function of target $P$ magnitude and the applied mask model. Interval bars represent dispersions at 90\% confidence level.}
                \label{fig:spr_tot_results}
        \end{center}
\end{figure}

We present in Figs.~\ref{fig:spr_k_results} and~\ref{fig:spr_tot_results} the results of $\mathrm{SPR}_{k}$ and $\mathrm{SPR}_{tot}$, respectively. The total SPR $\mathrm{(SPR}_{tot}$) was computed for all 50,000 sources of our working subset of targets, while the per contaminant SPR ($\mathrm{SPR}_{k}$) was computed for all $\sim3.25$ million stars located within a 10 pixel radius from those targets. Both plots show that the binary mask collects significantly less contaminant flux overall, and more particularly when the contaminant sources are located at more than 2 pixels distant from the targets. To give a rough idea, for about 80\% of the contaminant sources $\mathrm{SPR}_{k}$ is at least three times greater for the weighted masks. This result however is not surprising because gradient and Gaussian masks are typically larger to best fit the shape of the PSF. This is the reason why they typically give lower NSR, as shown in the previous section.

\subsection{Detectability of planet transits}
With both NSR and SPR determined, we are now capable of estimating the number $N^{good}_\mathrm{TCE}$ of target stars with sufficiently low NSR permitting the detection of eventual planets orbiting them. Tables~\ref{table:n_tce_good_a} and~\ref{table:n_tce_good_b} show the values for $N^{good}_\mathrm{TCE}$ for the case of an Earth-like planet orbiting a Sun-like star, respectively, for the scenarios of $\mathrm{SPR}_{tot}$ as given by Fig.~\ref{fig:spr_tot_results} (no background correction) and $\mathrm{SPR}_{tot}=0$ (perfect background correction).

\begin{figure}
        \begin{center}
                \includegraphics[height=12.0cm, width=\hsize]{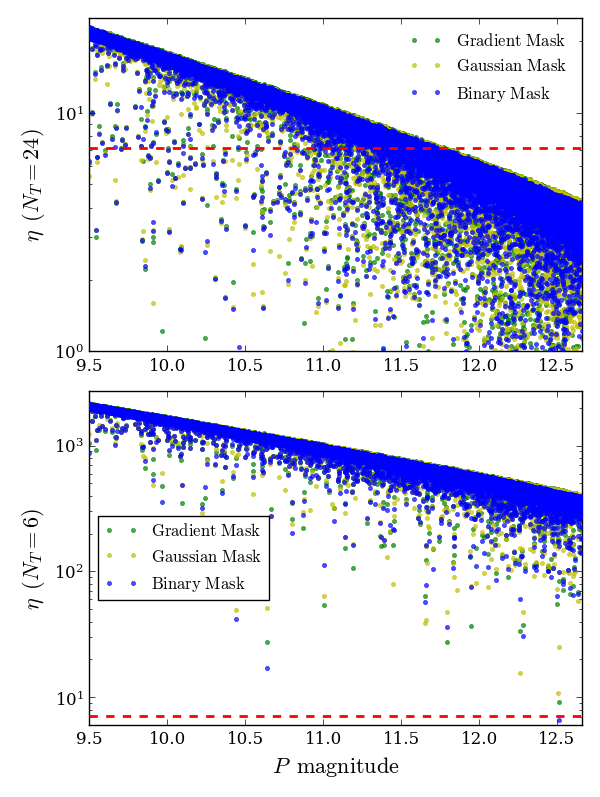}
                \caption{Scatter plot of the statistical significance $\eta$ (Eq.~\ref{eq:eta}) computed for 50,000 target stars, as a function of their respective $P$ magnitude and the applied aperture model. The red dashed line represents the threshold $\eta_{min}=7.1$. Values of $N^{good}_\mathrm{TCE}$ are provided in Tables~\ref{table:n_tce_good_a} and~\ref{table:n_tce_good_b}. \textbf{Top}: Statistics for an Earth-like planet with $\delta_p=84$ ppm, $t_d=13$h, $n_{tr}=3$, $\mathrm{SPR}_{tot}=0,$ and $N_T=24$. \textbf{Bottom}: Statistics for an Jupiter-like planet with $\delta_p=0.1$, $t_d=29.6$h, $n_{tr}=3$ ,$\mathrm{SPR}_{tot}=0,$ and $N_T=6$.}
                \label{fig:n_tce_good}
        \end{center}
\end{figure}

The results show that the advantage of weighted masks regarding NSR performance, which is up to $\sim7.5\%$ better with respect to the binary mask for the faintest and most numerous targets (see Fig.~\ref{fig:nsr_results}), does not translate into a proportionally better sensitivity in detecting true planet transits. Indeed, the mask with lowest NSR, called the gradient mask, provides only $\sim$0.8\% more chance of detecting Earth-like planets orbiting Sun-like stars at 1au. The difference between Gaussian and binary masks is even smaller, i.e. $\sim$0.4\%.  All three masks are equally capable of detecting Jupiter-like planets, no matter the number of telescopes observing the host star. To understand this, we need to compare Figs.~\ref{fig:nsr_results} and~\ref{fig:n_tce_good}. Taking the case of detecting Earth-like planets at about 1 au from Sun-like stars, our analyses show that the limiting magnitude\footnote{We note that this threshold is likely to be diminished by the presence of stellar activity in the noise \cite[see][]{Gilliland2011}.} for aperture photometry is of the order of $P\sim11.7$ ($V\sim12$ @6,000K) at $7.1\sigma$, $n_{tr}=3$ and $N_T=24$. Therefore, for most of the magnitude range ($11 \lesssim P \leq 12.66$) where the binary mask present the most degraded NSR performance with respect to the weighted masks, the latter do not provide any advantage in detecting such planets after all. Thus the small differences in $N^{good}_\mathrm{TCE}$ between binary and weighted masks, for the considered scenario, are consistent.

Correcting for the background results in an almost negligible impact ($\lesssim0.6\%$ increase) in the overall sensitivity of the apertures in detecting true planet transits. Also, it has no significant impact in the comparative basis analysis between the different aperture models. We stress however that inefficient background correction may significantly limit the accuracy with which planet transit depths can be determined.

Hence, from a planet transit finding perspective, designating an optimal solution for extracting photometry from the P5 stellar sample now becomes substantially less obvious. To this extent, looking at how each aperture performs in terms of false planet transit rejection may give us a hint about which is effectively the most appropriate choice.
\begin{table}
        \caption[]{Number $N^{good}_\mathrm{TCE}$ of target stars for which $\eta \geq \eta_{min}$, as a function of the number $N_T$ of telescopes observing them and the applied aperture model. We present below the case of an Earth-like planet with $\delta_p=84$ ppm, $t_d=13$h, $n_{tr}=3$ and $\mathrm{SPR}_{tot}$ given by the simulated values presented in Fig.~\ref{fig:spr_tot_results} (i.e. assuming no background correction). The values in this table were determined from our dataset of 50,000 target stars. The weighted values correspond to the effective $N^{good}_\mathrm{TCE}$, obtained by assuming uniform star distribution and a fractional field of view as given in Table~\ref{table:instrument_data}.}
        \label{table:n_tce_good_a}
        \centering
        \resizebox{\columnwidth}{!}{
                \begin{tabular}{c | c c c}
                        \hline
                        \noalign{\smallskip}
                        $N_T$ & Gradient mask & Gaussian mask & Binary mask \\ [0.5ex]
                        \hline 
                        \noalign{\smallskip}
                        24 & 19,063 (38.1\%) & 18,674 (37.3\%) & 18,201 (36.4\%) \\
                        \noalign{\smallskip}
                        18 & 15,105 (30.2\%) & 14,753 (29.5\%) & 14,469 (28.9\%)\\
                        \noalign{\smallskip}
                        12 & 10,629 (21.3\%) & 10,368 (20.7\%) & 10,202 (20.4\%) \\
                        \noalign{\smallskip}
                        6 & 5,528 (11.1\%) & 5,395 (10.8\%) & 5,357 (10.7\%) \\
                        \hline
                        \noalign{\smallskip}
                        \textbf{weighted} & \textbf{10,067 (20.1\%)} & \textbf{9,833 (19.7\%)} & \textbf{9,667 (19.3\%)} \\
                        \hline                                    
                \end{tabular}
        }
\end{table}
\begin{table}
        \caption[]{Same as Table~\ref{table:n_tce_good_a}, but for $\mathrm{SPR}_{tot} = 0$ (i.e. assuming a perfect background correction). A scatter plot of $\eta$, as a function of target $P$ magnitude, is illustrated in Fig.~\ref{fig:n_tce_good} for $N_T=24$.}
        \label{table:n_tce_good_b}
        \centering
        \resizebox{\columnwidth}{!}{
                \begin{tabular}{c | c c c}
                        \hline
                        \noalign{\smallskip}
                        $N_T$ & Gradient mask & Gaussian mask & Binary mask \\ [0.5ex]
                        \hline 
                        \noalign{\smallskip}
                        24 & 19,608 (39.2\%) & 19,319 (38.6\%) & 18,637 (37.3\%) \\
                        \noalign{\smallskip}
                        18 & 15,510 (31.0\%) & 15,264 (30.5\%) & 14,806 (29.6\%)\\
                        \noalign{\smallskip}
                        12 & 10,909 (21.8\%) & 10,701 (21.4\%) & 10,441 (20.9\%) \\
                        \noalign{\smallskip}
                        6 & 5,625 (11.2\%) & 5,527 (11.1\%) & 5,456 (10.9\%) \\
                        \hline
                        \noalign{\smallskip}
                        \textbf{weighted} & \textbf{10,318 (20.6\%)} & \textbf{10,141 (20.3\%)} & \textbf{9,884 (19.8\%)} \\
                        \hline                                    
                \end{tabular}
        }
\end{table}

\subsection{Sensitivity to background false transits}
We now compare the parameters $\mathrm{SPR}^{raw}_{k}$ (Eq.~\ref{eq:spr_raw}) and $\mathrm{SPR}^{app}_{k}$ (Eq.~\ref{eq:spr_app}) with $\mathrm{SPR}^{crit}_{k}$ (Eq.~\ref{eq:spr_critical}), to determine the number $N^{bad}_\mathrm{TCE}$ of contaminant stars with sufficiently high average flux to generate false positives. Two scenarios are considered herein: $N^{bad}_\mathrm{TCE}$ representing the number of contaminant sources for which $\mathrm{SPR}^{raw}_{k} \geq \mathrm{SPR}^{crit}_{k}$, which supposes no background correction in the photometry; and $N^{bad}_\mathrm{TCE}$ representing the number of contaminant sources for which $\mathrm{SPR}^{app}_{k} \geq \mathrm{SPR}^{crit}_{k}$, which supposes a perfect background correction in the photometry. In both cases, we define $\mathrm{SPR}^{crit}_{k}$ with $\delta_{back,k}=8.5\%$ ($\sim0.1$ mag), $t_d=4$h, $\eta=7.1,$ and $n_{tr}=3$. The chosen value for $\delta_{back,k}$ corresponds to the median depth of the sources in the {\it Kepler} Eclipsing Binary Catalogue (Third Revision)\footnote{\url{http://keplerebs.villanova.edu/}}, considering both primary (pdepth) and secondary (sdepth) depths together. The chosen value for $t_d$ corresponds to the median transit duration of the offset false positive sources listed in the Certified False Positive Table at NASA Exoplanet Archive\footnote{\url{https://exoplanetarchive.ipac.caltech.edu.}}. The transit duration values themselves were retrieved form the Threshold Crossing Events Table, by crossmatching the ID columns (KepID) from both tables.

\begin{figure}
        \begin{center}
                \includegraphics[height=11.0cm, width=\hsize]{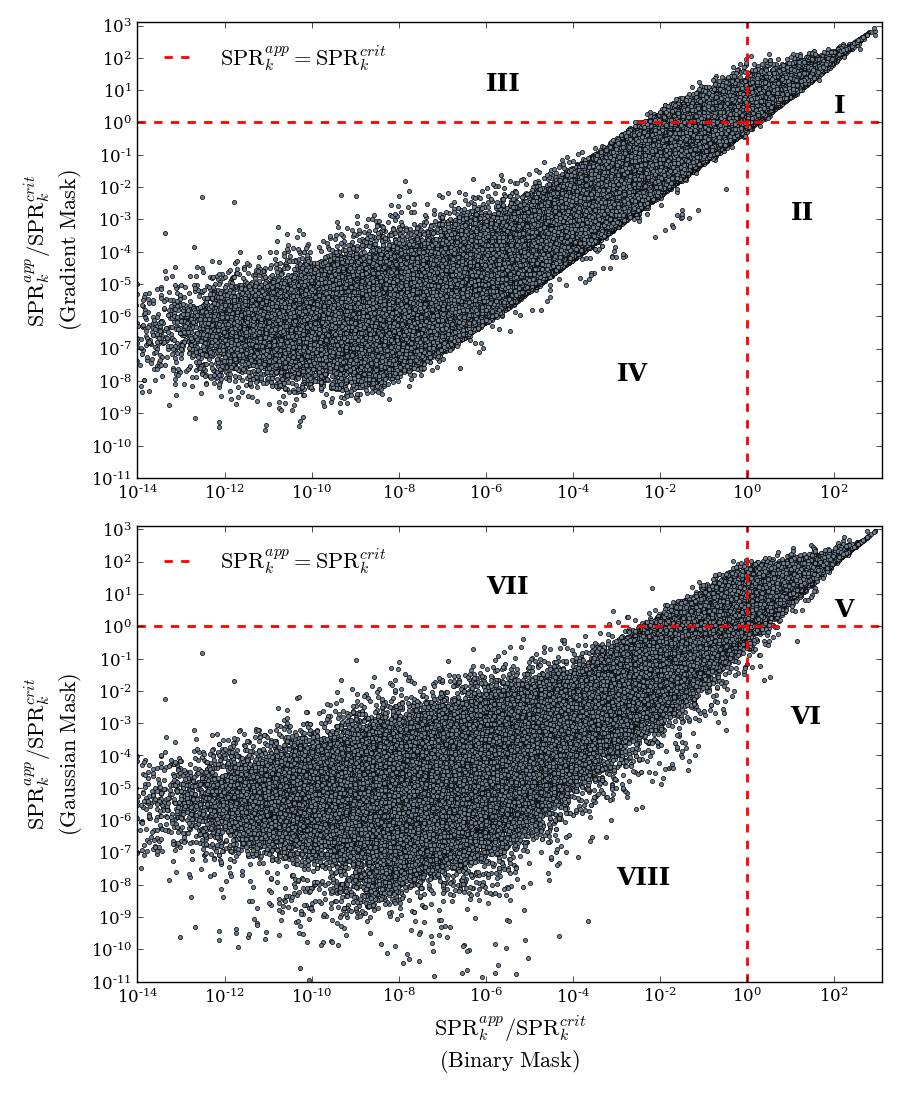}
                \caption{Scatter plot of $\mathrm{SPR}^{app}_{k}$ normalized by $\mathrm{SPR}^{crit}_{k}$, computed for $\sim3.25$ million contaminant stars. This illustration represents the particular case where $\mathrm{SPR}^{crit}_{k}$ is computed with $\delta_{back,k}\sim0.1$ mag; $N_T=12$; $n_{tr}=3$; $t_d=4$h. Values of $N^{bad}_\mathrm{TCE}$ are provided in Tables~\ref{table:n_tce_bad_a} and~\ref{table:n_tce_bad_b}. \textbf{Top}: Comparison between the values given by the gradient mask (vertical axis) and by the binary mask (horizontal axis). Region I: both masks exceed $\mathrm{SPR}^{crit}_{k}$. Region II: only the binary mask exceeds $\mathrm{SPR}^{crit}_{k}$. Region III: only the gradient mask exceeds $\mathrm{SPR}^{crit}_{k}$. Region IV: no mask exceeds $\mathrm{SPR}^{crit}_{k}$. \textbf{Bottom}: Comparison between the values given by the Gaussian mask (vertical axis) and the binary mask (horizontal axis). Regions V to VIII are analogous to I, II, III, and IV, respectively.}
                \label{fig:n_tce_bad}
        \end{center}
\end{figure}

Looking at the obtained results for $N^{bad}_\mathrm{TCE}$, which are presented in Tables~\ref{table:n_tce_bad_a} and~\ref{table:n_tce_bad_b}, the important thing to notice at first glance is the fact that all tested aperture models have, fortunately, an intrinsically very low (less than 5\%) overall sensitivity to detect mimicked planet transits caused by background eclipsing objects. In other words, these models are all insensitive to most of the potential false planet transits that may be produced by the contaminant sources in regions IV and VIII of Fig.~\ref{fig:n_tce_bad}. This is surely mostly because of the high enclosure energy of PLATO PSFs, but the optimization scheme applied to each aperture model, privileging low NSR, is also key in this context. Nevertheless, the results also clearly show that compared to the binary mask employing weighted masks substantially increases the predicted occurrence of events mimicking planet transits. The Gaussian mask is expected to deliver up to $\sim40\%$ higher $N^{bad}_\mathrm{TCE}$ than the binary mask, which is notably a huge discrepancy. The differences between gradient and binary masks are smaller, but still very significant: $N^{bad}_\mathrm{TCE}$ is up to $\sim20\%$ higher for the gradient mask. Either correcting for the background or not, these differences rest roughly the same, so background correction has no significant impact in the comparative basis analysis between the different aperture models. In absolute terms though, the results indicate that fully removing the background leads to an overall increase of more than 10\% in $N^{bad}_\mathrm{TCE}$, which is consistent with the analysis presented in Section~\ref{sec:background_corection}.

Overall, the obtained results for $N^{bad}_\mathrm{TCE}$, in comparison to those of $N^{good}_\mathrm{TCE}$ presented in the previous section, makes the scenario of choosing weighted masks become highly unfavoured even though that kind of mask provides better overall performance in terms of NSR. Still, it would be legitimate to ask whether the obtained values for $N^{bad}_\mathrm{TCE}$ are indeed significant in an absolute sense, since they represent less than 5\% of our full set of contaminant stars composed of $\sim$3.25 million sources. Properly answering this question requires carefully modelling the parameters $\delta_{back, k}$ and $t_d$ for the PLATO target fields, which is though beyond the scope of this paper. However, it is possible to obtain a rough idea of the occurrence of EBs ($N_\mathrm{beb}$) that could potentially result from the weighted values shown in Tables~\ref{table:n_tce_bad_a} and~\ref{table:n_tce_bad_b}. First, we need to consider that these values refer to about 20\% of the minimum number of expected targets for the P5 sample. Second, we may assume that the frequency of EBs ($F_\mathrm{eb}$) for the PLATO mission might be of the order of  1\%\footnote{\cite{Fressin2013} give $F_\mathrm{eb}=0.79\%$ for the {\it Kepler} mission. They defined it as being the fraction of EBs found by {\it Kepler}, including detached, semi-detached, and unclassified systems, divided by the number of {\it Kepler} targets.}. Accordingly, the expected occurrence of EBs at $7.1\sigma$, for the P5 sample could be approximately estimated with $N_\mathrm{beb} \sim 5 \times N^{bad}_\mathrm{TCE} \times 1\%$. From the weighted values presented in Tables~\ref{table:n_tce_bad_a} and~\ref{table:n_tce_bad_b}, that gives $1,600 \lesssim N_\mathrm{beb} \lesssim 2,500$ (all three tested aperture models comprised). This allows us to conclude that $N^{bad}_\mathrm{TCE}$ is thus not negligible. Moreover, considering that the total number of targets in the P5 sample is comparable to the total number of observed targets by the {\it Kepler} mission, we verified that our approximative estimate on the expected $N_\mathrm{beb}$, for the P5 sample, is very consistent to the statistics of background false positives of the {\it Kepler} mission. Indeed, the Certified False Positive Table on the NASA exoplanet archive gives at the present date 1,287 offset false positives out of 9,564 {\it Kepler} objects of interest. Such a consistency attests that our study is satisfactorily realistic. We stress however that accurate false positive estimates for the P5 sample cannot be provided by our study alone, in particular because it needs to be consolidated with PLATO's science exoplanet pipeline.  

\begin{figure}
        \begin{center}
                \includegraphics[height=18cm, width=7.5cm]{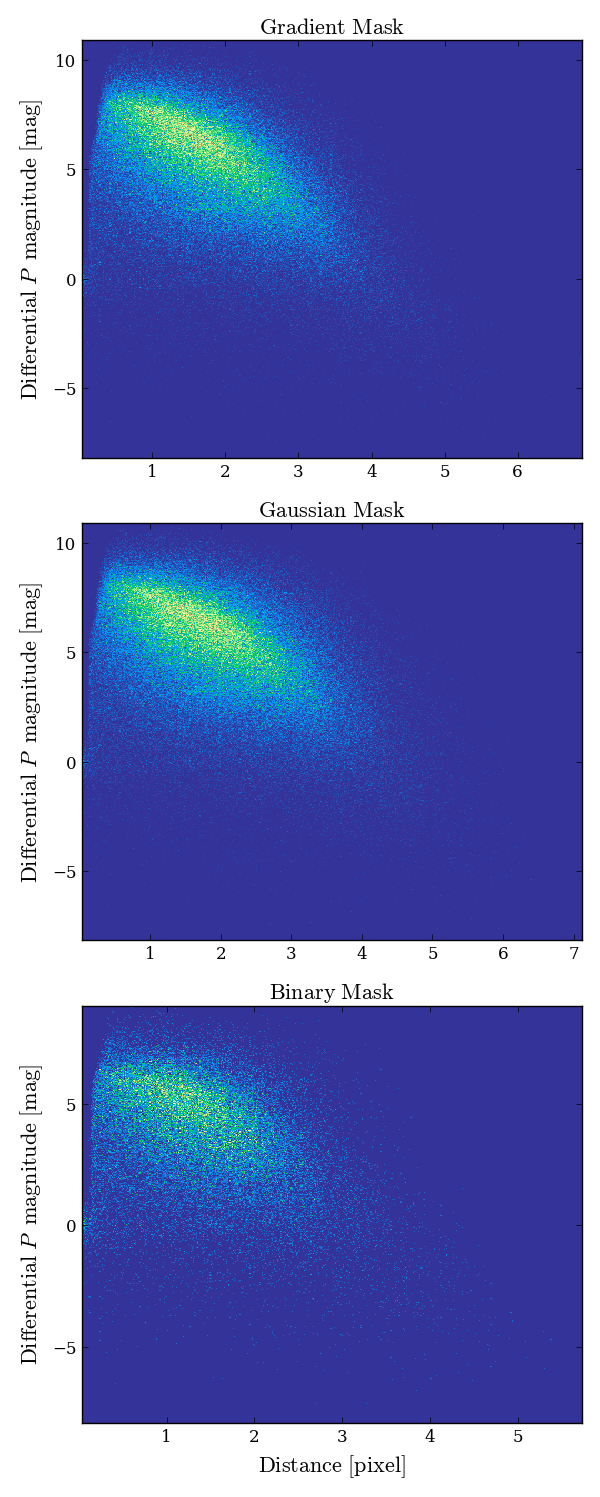}
                \caption{Two-dimensional histograms of the distribution of contaminant stars with $\mathrm{SPR}^{app}_{k} \geq \mathrm{SPR}^{crit}_{k}$, for gradient (top), Gaussian (centre), and binary (bottom) masks. The vertical axis indicates the differential $P$ magnitude between the contaminants and their respective targets, whereas the horizontal axis indicates the corresponding Euclidean distances. The parameters used to calculate $\mathrm{SPR}^{crit}_{k}$ are $\delta_{back,k}=0.8$ mag, $N_T=24$, $n_{tr}=3,$ and $t_d=4$h.}
                \label{fig:delta_mag_distance}
        \end{center}
\end{figure}

As a complement to the results presented in this section, Fig.~\ref{fig:delta_mag_distance} shows, for each aperture model, a two-dimensional histogram containing the distribution of contaminant stars having $\mathrm{SPR}^{app}_{k} \geq \mathrm{SPR}^{crit}_{k}$, as a function of the differential $P$ magnitude and the Euclidean distance between these sources and the corresponding targets. The parameters used to calculate $\mathrm{SPR}^{crit}_{k}$ were $\delta_{back,k}=0.8$ mag, $t_d=4$h, $N_T=24$, $\eta=7.1,$ and $n_{tr}=3$. This plot is of particular interest since it illustrates that the contaminant stars having sufficiently high average flux to produce background false positives are typically less than $\sim10$ mag brighter and located at less than $\sim4$ pixels away from the targets. Consequently, from the point of view of the distances, we verified that our approach of considering contaminant sources located at up to 10 pixels distant from the targets was largely enough for the purposes of this work. From the point of view of the differential magnitude, three important aspects need to be considered when interpreting the results.

First, we note that stars in our input catalogue are limited in magnitude to $P\sim21.1$. This means that for the faintest (and most numerous) P5 targets, for which $P$ magnitude is as high as 12.66, the maximum differential magnitude from their contaminants is therefore as small as $21.1 - 12.66 = 8.44$ mag, i.e. smaller than the limit of $\sim10$ mag suggested by the histograms. In contrast, P5 has targets as bright as 7.66 mag, so that the differential magnitude may be as high as $21.1 - 7.66 = 13.44$ mag. Hence, well above that limit.

Second, we notice in Fig.~\ref{fig:delta_mag_distance} some supposedly missing stars at distances near zero, in particular at differential magnitudes above 5 mag. We understand such an anomaly to be related to what we have already pointed out in Section~\ref{sec:stellar_sample} concerning bad estimates of the fluxes of stars fainter than $G\sim17$ in the DR2 catalogue. This issue is reported in \cite{Evans2018} and assumed to be caused by factors such as poor background estimation, observation taken in the proximity of bright sources, binarity, and crowding. In these conditions, the capability to isolate stars is therefore compromised. Taking into account that the most problematic cases were removed from the DR2 release according to the authors, the lack of stars in the above mentioned areas of Fig.~\ref{fig:delta_mag_distance} is justified. Yet scenarios of differential magnitude higher than 10 mag, at the same time that $\mathrm{SPR}^{app}_{k} \geq \mathrm{SPR}^{crit}_{k}$, should mostly occur at distances shorter than $\sim0.5$ pixel, where the occurrence of contaminant stars is substantially smaller than that at longer distances (see Fig.\ref{fig:target_contaminant_population}).

Third, the parameters used to build the histograms of Fig.~\ref{fig:delta_mag_distance} correspond in practice to a near worst case scenario in terms of the expected occurrences of false transits caused by background eclipsing objects. Indeed, it considers photometry perfectly corrected for the background; contaminants stars being observed by 24 cameras (maximum sensitivity to transit signatures); and contaminant stars generating background transit depths of 0.8 mag, which is significantly high. This means that the maximum differential magnitude is typically much smaller than 10 mag.

Taking into account all the above considerations, we conclude that Fig.~\ref{fig:delta_mag_distance} gives a sufficiently realistic and unbiased representation of distances and differential magnitudes of contaminant stars that are likely to cause background false planet transits, regardless of the limitation in maximum magnitude of our input catalogue. Furthermore, we note that the missing fraction ($\sim$0.01\%) of PSF energy in the images of Fig~\ref{fig:all_psfs} entails no significant impact in our analysis. This small fractional energy may be non-negligible uniquely in cases in which the differential magnitude between target and contaminant stars is $\lesssim-4$ mag. These are however extremely rare scenarios in our input stellar field, and thus statistically insignificant to our analysis. Indeed, less than 0.5\% of the contaminant sources in Fig.~\ref{fig:delta_mag_distance} have differential magnitude smaller than -2.6 mag. 

\begin{figure}
        \begin{center}
                \includegraphics[height=6.5cm, width=9cm]{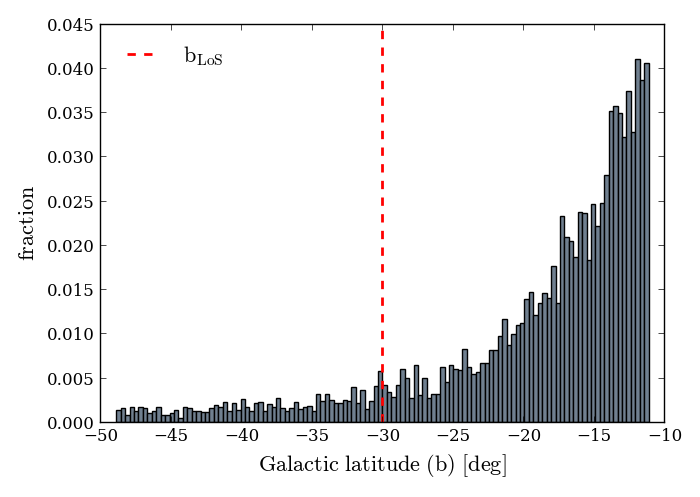}
                \caption{Fractional distribution of $N^{bad}_\mathrm{TCE}$, as a function of Galactic latitude (all three mask models comprised). This histogram \cite[cf.][]{Bryson2013} was built with contaminant sources that have Galactic longitude within the range $\mathrm{l_{LoS} \pm 1.5 \, [deg]}$. The red vertical line indicates the Galactic latitude $\mathrm{b_{LoS}}$ of $\mathrm{IF_{LoS}}$ (see IF coordinates in Table~\ref{table:los_coord}).}
                \label{fig:n_tce_bad_galactic_lat}
        \end{center}
\end{figure}

Ultimately, we extract the unique set of contaminant stars from all three histograms presented in Fig.~\ref{fig:delta_mag_distance}, and use it to build a histogram of the fractional distribution of contaminant stars having $\mathrm{SPR}^{app}_{k} \geq \mathrm{SPR}^{crit}_{k}$ (i.e. the fractional distribution of $N^{bad}_\mathrm{TCE}$) as a function of Galactic latitude. The resulting plot is shown in Fig.~\ref{fig:n_tce_bad_galactic_lat}. It suggests that the occurrences of false positives caused by background eclipsing stars might increase exponentially towards the Galactic plane, which is consistent with the distributions of offset transit signals presented in \cite{Bryson2013}. We note that since the distribution of stars within our IF privileges certain latitudes, because of its  its circular shape, we avoid propagating such bias to the data of Fig.~\ref{fig:n_tce_bad_galactic_lat} by considering contaminant sources within a sufficiently narrow Galactic longitude range $\mathrm{l_{LoS} \pm 1.5 \, [deg]}$ (see Table~\ref{table:los_coord}).
\begin{table}
        \caption[]{Number $N^{bad}_\mathrm{TCE}$ of contaminant stars for which $\mathrm{SPR}^{raw}_{k} \geq \mathrm{SPR}^{crit}_{k}$, which supposes photometry with no background correction, as a function of the number $N_T$ of telescopes observing the host star and the aperture model. The presented values were determined from our dataset of $\sim3.25$ million contaminant stars. The $\mathrm{SPR}^{crit}_{k}$ was computed with $\delta_{back,k}\sim0.1$ mag, $t_d=4$h, $\eta=7.1,$ and $n_{tr}=3$. The roman numerals correspond to the areas indicated in Fig.~\ref{fig:n_tce_bad}. The percentiles indicate the amount of deviation of the values from weighted masks with respect to those from binary mask. The weighted values in the lower row correspond to the effective $N^{bad}_\mathrm{TCE}$, obtained by assuming uniform star distribution and a fractional field of view as given in Table~\ref{table:instrument_data}.}
        \label{table:n_tce_bad_a}
        \centering
        \resizebox{\columnwidth}{!}{
                \begin{tabular}{c | c c c}
                \hline
                \noalign{\smallskip}
                        $N_T$ & Binary mask & Gradient mask & Gaussian mask \\ [0.5ex]
                              & $\mathrm{(I + II)=(V + VI)}$    & (I + III)     & (V + VII)     \\
                        \hline 
                        \noalign{\smallskip}
                        24 & 40,135 & 48,005 & 55,520 \\
                        \noalign{\smallskip}
                        18 & 36,835 & 43,690 & 50,785 \\
                        \noalign{\smallskip}
                        12 & 32,830 & 38,485 & 44,565 \\
                        \noalign{\smallskip}
                        6  & 26,545 & 31,050 & 35,995 \\
                        \hline
                        \noalign{\smallskip}
                        \textbf{weighted} & \textbf{31,591} & \textbf{37,178 (+17.7\%)} & \textbf{43,073 (+36.3\%)} \\
                        \hline                                    
                \end{tabular}
        }
\end{table}
\begin{table}
        \caption[]{Same as Table~\ref{table:n_tce_bad_a}, but now representing the contaminant stars for which $\mathrm{SPR}^{app}_{k} \geq \mathrm{SPR}^{crit}_{k}$, which supposes photometry with perfect background correction.}
        \label{table:n_tce_bad_b}
        \centering
        \resizebox{\columnwidth}{!}{
                \begin{tabular}{c | c c c}
                        \hline
                        \noalign{\smallskip}
                        $N_T$ & Binary mask & Gradient mask & Gaussian mask \\ [0.5ex]
                              & $\mathrm{(I + II)=(V + VI)}$    & (I + III)     & (V + VII)     \\
                        \hline 
                        \noalign{\smallskip}
                        24 & 45,180 & 54,720 & 63,555 \\
                        \noalign{\smallskip}
                        18 & 41,575 & 50,185 & 58,540 \\
                        \noalign{\smallskip}
                        12 & 37,055 & 44,380 & 51,885 \\
                        \noalign{\smallskip}
                        6  & 30,185 & 35,820 & 41,570 \\
                        \hline
                        \noalign{\smallskip}
                        \textbf{weighted} & \textbf{35,731} & \textbf{42,774 (+19.7\%)} & \textbf{49,814 (+39.4\%)} \\
                        \hline                                    
                \end{tabular}
        }
\end{table}

\subsection{Implementation constraints} \label{sec:implementation}

\subsubsection{Updating the masks on board} \label{sec:updating_masks}
As explained earlier in Section~\ref{sec:psf}, the pixels of PLATO detectors are relatively broad compared to the size of the PSFs. During observations, this causes aperture photometry to be sensitive to the long-term star position drift occurring on the focal plane. For PLATO, this effect is expected to be caused notably by the orbital differential velocity aberration\footnote{\url{http://www.stsci.edu/hst/fgs/documents/datahandbook}} and the thermo-elastic distortion from the optical bench, and might be as large as 1.3 pixel over three months. Consequently, mask-target assignments performed during each calibration phase become, soon or later, no longer optimal, since the flux distributions of the targets significantly change as these move across the pixels. Therefore, the NSR of the resulting light curves substantially increases.

To compensate for this effect, the proposed solution consists in tracking the targets by updating the placement of their apertures on board, as explained in \cite{Samadi2019}. This will involve both ground and flight segments of the mission, as the apertures will first be computed on the ground and then transmitted to the spacecraft. Both the criteria and timescale on which such actions will be performed are yet to be defined.

\subsubsection{Uploading the masks on board} \label{sec:upload_masks}
The performance results presented in this paper were obtained by assigning apertures for each target individually, following the computation schemes presented in Section~\ref{sec:aperture_models}. Taking the case of the gradient mask, which provides the lowest values of NSR, having such performance on board requires a unique mask shape per target to be uploaded to the flight software. This demands, in turn, prohibitive telemetry and time resources. In addition, as explained in Section~\ref{sec:updating_masks}, the masks will have to be regularly updated in flight to compensate for long-term star position drift, thus making the employment of gradient masks unfeasible.

For the Gaussian mask the outlook is not much more favourable, as this solution would require a massive set of widths (practically one per target) to guarantee the NSR performance results presented in Fig.~\ref{fig:nsr_results}. Otherwise, we could in principle take advantage of the fact that the Gaussian mask has an analytical form -- with small number of parameters -- to apply simplification schemes to avoid the need for having one particular mask per target. For instance, a feasible approach would consist of employing polynomial surfaces or fixed widths to cover the multiple combination scenarios in terms of magnitude and intra-pixel location of the targets. Nevertheless, that would inevitably reduce the overall NSR performance, which is the major benefit of using weighted masks.

\begin{figure}
        \begin{center}
                \includegraphics[height=9.0cm, width=\hsize]{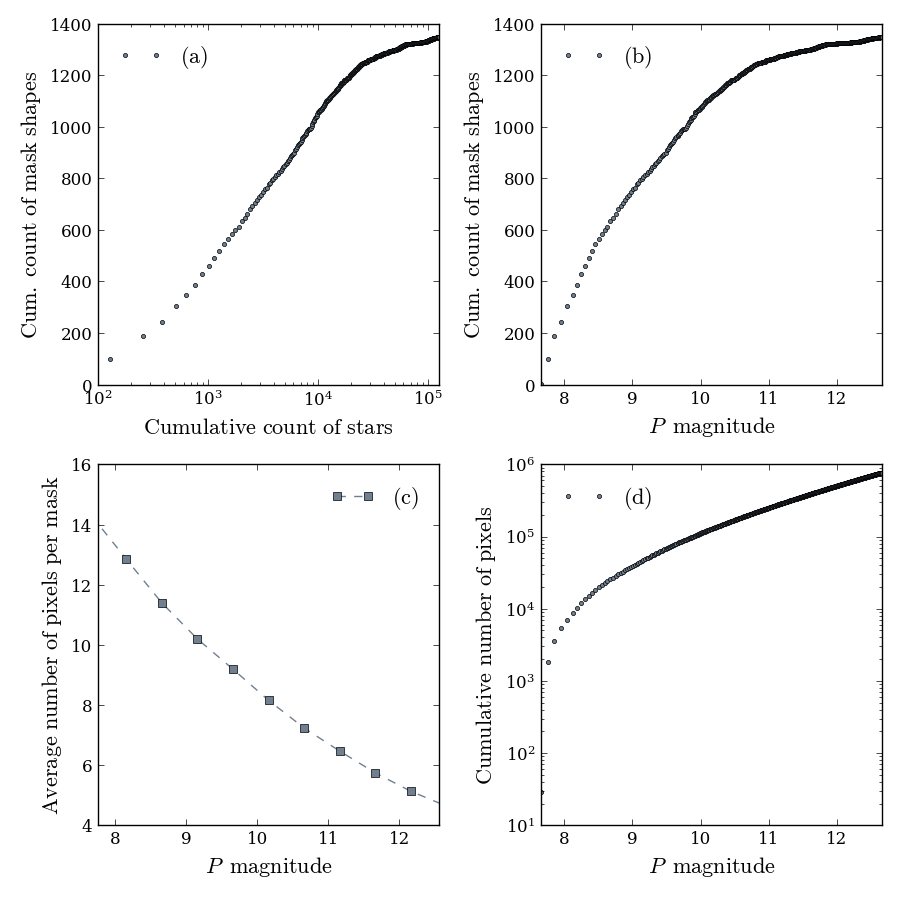}
                \caption{Statistics on the morphology of binary masks. The above results are based on all $\sim127$ thousand target stars within IF (see Section~\ref{sec:stellar_sample}). \textbf{(a)}: Cumulative count of unique binary mask shapes as a function of the cumulative count of target stars. \textbf{(b)}: Cumulative count of unique binary mask shapes as a function of target $P$ magnitude. \textbf{(c)}: Average number of pixels composing the binary masks as a function of target $P$ magnitude. \textbf{(d)}: Cumulative number of pixels composing the binary masks as a function of target $P$ magnitude.}
                \label{fig:binary_mask_libraries}
        \end{center}
\end{figure}

The binary mask, in turn, provides a virtually unbeatable capacity for compressing combinations of mask shapes without loss of performance. We can visualize this by looking at the data concerning the morphology of binary masks provided in Fig.~\ref{fig:binary_mask_libraries}a. These data give the accumulated unique combinations of binary mask shapes computed from the set of binary masks used to extract photometry from all $\sim127$ thousand target stars in our adopted IF. We verify that the unique set saturates to about only 1,350 mask shapes, thereby giving a compression factor of almost 99\%. This represents another significant advantage of employing the binary mask, since no weighted mask is actually capable of providing such compression capabilities while keeping the original performance of the full set of masks unchanged.

We note that the statistics on the number of binary mask shapes and pixels, presented in Fig.~\ref{fig:binary_mask_libraries}, are valid for non-saturated stars. It implies that only one mask is attributed to each target, and each mask is limited in size by the ($6\times6$) shape of an imagette. This is a fundamental assumption for the study presented herein. In the current instrument design, PLATO detectors are expected to exhibit saturation at pixels observing stars brighter than $P \sim 8.16 \pm 0.5$ (i.e. $V \sim 8.5 \pm 0.5$ @6,000K) after a 25s exposure (normal cameras). The exact saturation limit depends on the location of the star in the CCD and where its barycentre falls within a pixel. The brightest stars in our study are thus at the very lower bound of this broad saturation threshold. In the context of the PLATO science pipeline, the photometry of saturated stars will be extracted exclusively from the ground on the basis of extended imagettes, that is, nominal imagettes appropriately extended such as to capture the charges spilt along the CCD columns from the saturation.

\section{Conclusions and discussions} \label{sec:conclusion}
Light curves will be produced in flight for potentially more than 250,000 PLATO targets (the P5 stellar sample) by employing aperture photometry. To maximize the scientific exploitability of the resulting data, an appropriate aperture model needs to be determined. Aiming to fulfil this objective, we presented in this paper a detailed photometric performance analysis based on three different strategies: a weighted aperture providing global minimum NSR (gradient mask) obtained through a novel direct calculation method; a weighted Gaussian aperture giving suboptimal NSR; and a narrower binary aperture to reduce the impact of contamination. Each aperture model was applied to 50,000 synthetic imagettes containing a realistic stellar distribution with more than three million sources, extracted from the {\it Gaia} DR2 catalogue. The stellar population was obtained from one of the expected long-pointing fields for the mission, namely the southern PLATO field.

For a more appropriate estimate of stellar fluxes reaching the instrument's detectors, we established a synthetic PLATO $P$ photometric passband derived from the spectral response of the instrument and calibrated in the VEGAMAG system. This allows us to avoid the inconvenience of having colour dependency when estimating stellar fluxes from $V$ magnitudes. The photometric relationships $V-P$ and $G-P$ are included. In addition, we used a zodiacal light semi-analytical model from the literature to derive an expression for estimating the intensities of scattered background light entering the PLATO cameras.

To determine the optimal aperture model for extracting photometry from the P5 targets, we adopted an innovative criterion that is based on two science metrics: a simulated number of target stars for which a planet orbiting it would be detected, denoted as $N^{good}_\mathrm{TCE}$ (to be maximized); and a simulated number of contaminant stars that are sufficiently bright to generate background false positives when eclipsed, denoted as $N^{bad}_\mathrm{TCE}$ (to be minimized). Both metrics depend on NSR, SPR, and simulated frequency of TCEs at 7.1$\sigma$; they allow a direct evaluation of the scientific performance of apertures in detecting true and false planet transit signatures. The {\it Kepler} and TESS missions adopt, analogous to our stellar pollution (SPR), the crowding metric $r$ \cite[]{Batalha2010} and the dilution parameter $D$ \cite[]{Sullivan2015}, respectively, to quantitatively distinguish photometric fluxes originating from targets and other sources. However, these are instrumental level parameters and are not taken into account for choosing their apertures.

From our results we conclude that, compared to the binary mask, weighted masks (gradient and Gaussian) best fit the instrumental PSF at pixel resolution, thus providing lower NSR in general, but their larger wings inevitably encompass more fractional flux from contaminant stars. From a science perspective, all three mask models present comparable overall efficiency in detecting legitimate planet transits, but the binary mask is substantially (up to $\sim$30\%) less likely to produce background false positives with respect to the weighted masks. These results led us to select the binary mask as the optimal solution for extracting photometry in flight from P5 targets, since this provides the best compromise between maximizing $N^{good}_\mathrm{TCE}$ and minimizing $N^{bad}_\mathrm{TCE}$. Besides, this mask model offers a significant implementation advantage, since it requires a relatively small number of unique mask shapes to extract photometry from a large set of stars; we found that about 1,350 unique binary masks are sufficient to extract optimal photometry from $\sim127$k targets.

Our approach currently represents a consistent contribution to the science of exoplanet searches. It confirms that the ordinary concept adopted in the literature for finding apertures, which typically relies on noise minimization for maximum transit detection without directly taking into account the impact from false positives, is not necessarily the best strategy. This statement was initially raised as a hypothesis earlier in this paper, and our results confirm that it holds for the PLATO P5 sample. Indeed, the conventional approach would suggest the use of weighted masks instead of the binary mask.

Beyond the P5 sample, the weighted masks may be exploited as additional photometry extraction methods for the targets whose light curves will be produced  from the ground from imagettes. Compared to more complex methods based on PSF fitting photometry \cite[e.g.][]{Libralato2015, Nardiello2016}, our gradient and Gaussian masks are much simpler and faster to calculate. They might be suitable for not too crowded fields or in situations in which the existence of contaminants may not be too critical (e.g. for asteroseismology targets). We note however that these masks adapt their size to the presence of contaminant stars. This is possible since our expression for the NSR (Eq.~\ref{eq:nsr_cadence}) takes into account the fluxes coming from contaminant companions, so whenever their signals are sufficiently strong compared to those of the targets the masks are reduced in width to keep NSR as low as possible. Moreover, our weighted masks can be implemented with ease in both {\it Kepler} and TESS data processing pipelines, so their usage is not limited to PLATO targets. We expect that the ensemble of results and discussions derived from this work might be particularly useful during the next steps of the preparation phases of the PLATO mission, in particular for the definition of algorithms in the exoplanet validation pipeline, for the construction of the PIC, and later on for the selection of targets.

Finally, despite the relevant contributions of the present study towards minimizing the frequency of background false positives in the P5 sample, a particular concern might still arise with regard the potential difficulties in properly identifying, based on the light curves alone, the false positives from the P5 detections. We highlight however that for an observation scenario covering two long pointing fields the P5 photometry includes, in addition to light curves, a dedicated data share comprising more than 9,000 imagettes -- with 25 seconds cadence -- and COBs for 5\% of the targets \cite[see][]{ESA2017_RBD}. Allocating these resources to the P5 targets is expected to be flexible enough so that they can be employed following the principle of an alert mode, for example whenever transit signals are detected in the light curves available on the ground. Therefore, the P5 sample will be composed of a photometry extraction method (binary masks) that is intrinsically insensitive to detect most of the potential background false positives, plus a non-negligible number of imagettes and COBs that can be strategically allocated to targets of interest. Overall, that should be enough to identify properly a substantial fraction of the TCEs, which will be dominated by short period transits, in the P5 sample. Aside from that, the PLATO data processing team is currently studying the feasibility and effectiveness of applying imagette-independent methods for identifying background false positives from the P5 detections.

\begin{acknowledgements}
        This work has benefited from financial support by the French National Centre for Space Studies (CNES), the Paris Observatory -- PSL and the Brazilian National Council for Scientific and Technological Development (CNPq). This work has made use of data from the European Space Agency (ESA) mission {\it Gaia} (\url{https://www.cosmos.esa.int/gaia}), processed by the {\it Gaia} Data Processing and Analysis Consortium (DPAC, \url{https://www.cosmos.esa.int/web/gaia/dpac/consortium}). Funding for the DPAC has been provided by national institutions, in particular the institutions participating in the {\it Gaia} Multilateral Agreement. This research was achieved using the POLLUX database\footnote{\url{http://pollux.graal.univ-montp2.fr}} operated at LUPM  (Université Montpellier - CNRS, France with the support of the PNPS and INSU. This research uses solar spectrum data from the National Renewable Energy Laboratory, operated for the U.S. Department of Energy (DOE) by the Alliance for Sustainable Energy, LLC ("Alliance"). This research made use of Astropy,\footnote{\url{http://www.astropy.org}} a community-developed core Python package for Astronomy \cite[]{AstropyCollaboration2013}; Scipy,\footnote{\url{http://www.scipy.org/}} a Python-based ecosystem of open-source software for mathematics, science, and engineering; and SunPy\footnote{\url{http://www.sunpy.org}}, an open-source and free community-developed solar data analysis package written in Python \cite[]{SunPyCommunity2015}. The authors express their deep gratitude to the anonymous referee for her/his constructive remarks and suggestion; give special thanks of gratitude to Valerio Nascimbeni for his courtesy in providing the illustration of Fig.~\ref{fig:plato_fields}; and gratefully acknowledge the helpful suggestions by David Brown and Beno{\^{i}}t Mosser and the language editing by Amy Mednick.
\end{acknowledgements}

\bibliographystyle{aa} 
\bibliography{bibliography}

\end{document}